\documentclass[10pt,journal,compsoc]{IEEEtran}
\usepackage{ifpdf}

\ifCLASSOPTIONcompsoc
  \usepackage[nocompress]{cite}
\else
  \usepackage{cite}
\fi

\ifCLASSINFOpdf
   \usepackage[pdftex]{graphicx}
   \graphicspath{{../pdf/}{../jpeg/}}
   \DeclareGraphicsExtensions{.pdf,.jpeg,.png}
\else

\fi

\usepackage{amsmath}

\usepackage{algorithmic}

\usepackage{array}

\usepackage{fixltx2e}

\usepackage{stfloats}
\usepackage{url}

\usepackage{booktabs} 

\usepackage{multirow}

\usepackage{color}

\usepackage{threeparttable}

\usepackage{makecell}

\usepackage{amssymb}

\usepackage{bbding}

\begin{document}
%
\title{Security and Privacy for Healthcare Blockchains}
\author{Rui~Zhang,
        Rui~Xue,
        and~Ling~Liu,~\IEEEmembership{Fellow,~IEEE}
\IEEEcompsocitemizethanks{\IEEEcompsocthanksitem R. Zhang is with State Key Laboratory of Information Security, Institute of Information Engineering, Chinese Academy of Sciences, Beijing, 100093 China, and the College of Computing, Georgia Institute of Technology, Atlanta, GA, 30332 USA.\protect\\
E-mail: zhangrui@iie.ac.cn
\IEEEcompsocthanksitem R. Xue is with State Key Laboratory of Information Security, Institute of Information Engineering, Chinese Academy of Sciences, Beijing, 100093 China, and the School of Cyber Security, University of Chinese Academy of Sciences, Beijing, 100049 China.\protect\\
E-mail: xuerui@iie.ac.cn
\IEEEcompsocthanksitem L. Liu is with the College of Computing, Georgia Institute of Technology, Atlanta, GA, 30332 USA\protect\\ E-mail:lingliu@cc.gatech.edu.
\IEEEcompsocthanksitem This paper has been accepted for publication in IEEE Transactions on Services Computing. DOI: 10.1109/TSC.2021.3085913
}
}

\IEEEpubid{\copyright 2021 IEEE. Personal use is permitted, but republication/redistribution requires IEEE permission. See https://www.ieee.org/publications/rights/index.html for more information.}


\IEEEtitleabstractindextext{%
\begin{abstract}
Healthcare blockchains provide an innovative way to store healthcare information, execute healthcare transactions, and build trust for healthcare data sharing and data integration in a decentralized open healthcare network environment.
Although the healthcare blockchain technology has attracted broad interests and attention in industry, government and academia, the security and privacy concerns remain the focus of debate when deploying blockchains for information sharing in the healthcare sector from business operation to research collaboration.
This paper focuses on the security and privacy requirements for medical data sharing using blockchain, and provides a comprehensive analysis of the security and privacy risks and requirements, accompanied by technical solution techniques and strategies. First, we discuss the security and privacy requirements and attributes required for electronic medical data sharing by deploying the healthcare blockchain. Second, we categorize existing efforts into three reference blockchain usage scenarios for electronic medical data sharing, and discuss the technologies for implementing these security and privacy properties in the three categories of usage scenarios for healthcare blockchain, such as anonymous signatures, attribute-based encryption, zero-knowledge proofs, verification techniques for smart contract security. Finally, we discuss other potential blockchain application scenarios in healthcare sector. We conjecture that this survey will help healthcare professionals, decision makers, and healthcare service developers to gain technical and intuitive insights into the security and privacy of healthcare blockchains in terms of concepts, risks, requirements, development and deployment technologies and systems.
\end{abstract}

}

\maketitle

\IEEEdisplaynontitleabstractindextext

\IEEEpeerreviewmaketitle

\IEEEraisesectionheading{\section{Introduction}\label{sec:introduction}}
\IEEEPARstart{I}{n} the healthcare sector, key patient data and information are created by and maintained across different healthcare organizations and departments using possibly heterogeneous healthcare systems. As a result, health care providers cannot easily access critical data when they need it for delivering quality care services, including diagnosis, treatment decision making and recommendations. Furthermore, when medical treatments are complex and require multiple healthcare professionals from different healthcare providers or related organizations to access the electronic medical records (EMRs) of the patients under treatments, there lacks efficient EMR sharing services and systems to provide flexibility, control and administrative management of timely and secure access to relevant EMR data. As the healthcare provisioning is embracing the Internet of Things (IoT) technologies, many healthcare and medical facility providers are equipped with smart gadgets, such as handheld devices, laptop or desktop computers with a variety of healthcare applications. However, it remains to be a grand challenge for healthcare providers, when needed, to seamlessly collect, analyze and exchange EMR data on demand with proper security and privacy guarantee, which, as recognized by many~\cite{Engelhardt:2017:HHC}, may greatly hinder the quality of real-time healthcare services, and at the same time, it may require the patients with or without health insurance to spend more medical expenses to request their relevant historical healthcare records to be sent to the current caregivers. These real-time healthcare data sharing scenarios show the urgent demand for a timely and secure access and trusted and possibly decentralized EMR data sharing infrastructure in the healthcare sector to provide smooth, transparent, cost-effective, and easy to operate EMR data sharing on demand and in real-time.\\

\noindent \textbf{Motivation of Blockchain.}
A healthcare information system handles sensitive patient information and requires fast and secure access to healthcare data to provide better healthcare. The blockchain as a secure distributed ledger provides a potential solution for cross-system sharing of medical information with a number of advantages:

\noindent \textbf{(1) Chronological patient records.} On the blockchain, all of the patient's medical records, including outpatient, inpatient, and wearable device tests, are automatically chronologically ordered for patient and physician review and provide step-by-step medical care.

\noindent \textbf{(2) Data verification without a central authority.} The blockchain allows for public validation statements where the network agrees to implement consensus and does not require any central authority to participate in and manage the entire network.

\noindent \textbf{(3) Secure storage with integrity guarantee.} With blockchain, medical data can be stored on the hash chain without being tampered with, even when a small portion of nodes on the distributed the healthcare block chain are compromised by an adversary, the entire healthcare blockchain system can remain to function without interruption.

\noindent \textbf{(4) Data sharing with efficient interoperability.} Blockchain enables efficient interoperability by providing reliable and scalable storage and access to medical data across institutional and administrative boundaries using smart contracts with easy-to-use application interfaces (APIs). The healthcare blockchain holds high potential to reduce and possibly eliminate the current costs, time-delay and administrative burdens that arise for medical data sharing and coordination across autonomous healthcare organizations.

Hence, the blockchain can serve as a decentralized platform for collecting, storing, and sharing medical records in a scalable and secure manner.
As the adoption rate of the blockchain continues to increase in digital currency, financial industry and business operations, since the birth of Bitcoin~\cite{Nakamoto:2008:Bitcoin}, the blockchain has entered the medical industry. Even at its infancy, the technology is being developed, and deployed by players in the healthcare ecosystem.\\

\noindent \textbf{Scope and Contributions.}
Although employing the blockchain technology to healthcare for efficient sharing of medical data is innovative and disruptive, it is not a panacea for addressing the security and privacy requirements and risks of EMR data. Instead, we argue that in-depth and comprehensive understanding of the challenges and corresponding technologies of security and privacy in healthcare blockchain will provide guidance and technological evolution for the next generation of healthcare ecosystem.

Although numerous academic and industrial projects and products in healthcare blockchain, most existing products have focused only on the secure storage and access control of medical records using the security attributes inherent in the blockchain itself. Some cryptographic methods are proposed to implement additional security attributes, such as the confidentiality of medical records. However, there are little dedicated efforts to provide a comprehensive analysis and a detailed explanation of the security and privacy in the healthcare blockchain, which is arguably important because medical information stored in the healthcare blockchain is highly sensitive to both patients and care providers in terms of privacy and security of medical information.

The survey conducts a comprehensive analysis and in-depth review in terms of the security and privacy in the healthcare blockchain systems and applications. We first discuss the notions of healthcare blockchain and its security and privacy requirements. We then describe a set of security techniques, especially cryptographic solutions,  corresponding to three healthcare application scenarios, for supporting the additional security goals. Finally, we discuss other potential blockchain applications in the healthcare sector. We believe that as blockchain technology becomes more mature and to be applied in the healthcare sector, gaining insight into the security and privacy of the healthcare blockchain will help to address the underlying vulnerabilities in existing healthcare blockchain systems and provide strong defense technologies and technological innovations to ensure the security and privacy of the healthcare blockchain.

This review article aims to serve dual goals. First, it will help non-security experts better understand the security and privacy requirements and risks of the healthcare blockchain. Second, it will embrace healthcare professionals, researchers and practitioners with the leading security and privacy technologies for developing and deploying the healthcare blockchain. It will also discuss the security and privacy properties important to the healthcare blockchain, and present corresponding solutions to achieve these security goals, and point out open problems and challenges. Furthermore, it may also guide medical scientists and engineers to develop appropriate healthcare blockchain systems and techniques based on specific application scenarios.

\section{Multifacets of EMRs}
\label{sec:multifacets-emr}

\subsection{Concepts of EMRs}
\label{subsec:concepts}
First, we need to introduce some definitions of healthcare blockchain that are mentioned in this paper. For a more detailed description of these concepts, please refer to \cite{Rui:2010:SMR}.

\textbf{Care Delivery Organization (CDO)} is an organization that can provide medical services including hospitals, medical centers, etc.

\textbf{Electronic Medical Records (EMRs)} are legal medical records that are generated when patients are inpatient and outpatient at a CDO. EMR is created and used by healthcare practitioners to accurately record the healthcare services that the patient receives within CDO. EMR is maintained by the CDO that provides appropriate medical services to patients.

\textbf{Electronic Health Record (EHR)} is a subset of the EMRs that created and maintained by each CDO. Based on the ISO/TS 18308~\cite{ISO/TS:18308}, EHR refers to EMRs that can be shared across CDOs via patient consent to provide patient with quality and convenient healthcare services. An EHR is usually jointly managed and controlled by the patient and the CDO, that is, it should be agreed or authorized by the patient that who and when his or her EHR can be accessed.

\textbf{Personal Health Record (PHR)} is usually a health record stored and maintained by an individual that includes individuals collecting data from many sources, i.e., data monitored by the wearable device at home, EMR, and EHR. A complete history of PHRs can help healthcare provider provide more accurate treatment for individuals.

\textbf{Healthcare Blockchain} is a blockchain used to store and/or share medical data. CDOs or individuals store all or part of the medical information on the blockchain by establishing a blockchain owned by themselves or purchasing blockchain services from a blockchain platform provider and utilize the inherent characteristics of the blockchain such as non-tamperable and distributed to realize secure storage and sharing of medical information.

It is worth noting that the healthcare blockchain has very different privacy and security requirements from those of the financial healthcare payment based on blockchain. The former is a distributed database where EMR data and related information are stored and accessed, and the latter is a blockchain based payment system that allows users to pay for medical bills in cryptocurrency such as using Bitcoin~\cite{Nakamoto:2008:Bitcoin} or Ethereum~\cite{Ethereum}. Due to the space constraint, this paper focuses on the security and privacy of healthcare blockchain, and the healthcare payment blockchain specific security and privacy issues, such as double-spending attack and wallet security, are beyond the scope of this paper.

\subsection{Security and Privacy Concerns of EMRs}
\label{subsec:concerns}
When medical records change from paper to electronic medical records to bring convenience to doctors and patients, it also poses security and privacy challenges. Meingast et al.~\cite{Meingast:2006:SPI} summarized the security and privacy issues in healthcare information technology and pointed out that in order to address the security and privacy challenges of healthcare information system and the cross-institutional transmission and sharing of electronic medical records, the following questions need to be considered and answered first:

\textbf{(1) Who owns the medical record?} Who has the right to create, delete and edit medical records? Do individual patients have ownership of medical records collected by themselves? Do their doctors or insurance companies own the medical records? In addition, when a medical record is given to a third party, does he or she have the same rights as the data owner, or are some of their rights restricted?

\textbf{(2) Who can read/write a patient's medical record?} Who has the right to write a medical record for a patient? Can medical records be modified when data or diagnostic errors occur? Who has the right to modify the medical records that already have been written? Who has the right to read a medical record? What portion of the medical record should be shared with different users?

\textbf{(3) Should medical data be leaked to persons other than the previously authorized users without the patient's consent?} Note that, authorized users refer to users who are granted access right to specific medical data. When a special situation occurs, for example, when the patient suddenly falls into a coma, is it possible to disclose medical data to someone other than the owner and previously authorized users without the patient's consent? To whom should this medical data be disclosed? Who has the right to do this?

\textbf{(4) Where should the electronic medical records be stored?} For electronic medical records, should the data be stored in a local database that can be connected to each other via a network, or should they be stored in a central database accessible to all users? What type of data storage best meets the security and privacy requirements of electronic medical records?\\

For answering the question of ``who owns a medical record?", we should first clarify the notions of ``creator", ``manager", and ``owner" of medical records. \textbf{``Creator"} is the person who creates a medical record and generates the data and content on it. In a healthcare information system, the attending physician or laboratory staff is the creator of the patient's medical data. \textbf{``Manager"} is an individual or entity responsible for managing, monitoring, and protecting information. In a healthcare system, both the patient and system administrator are managers, where patient decides who can access his medical records, and the system administrator ensures the secure operation of a healthcare system where the medical records are created, stored, and accessed. \textbf{``Owner"} means the person or entity who has the legal or rightful title to something. In a healthcare system, although medical records contain patient's personal health information and were created for them, the owners of electronic health records are not necessarily the patients.

Under federal and state law, patients have legal privacy, security and accuracy rights related to their health information. However, once that information is captured and documented in written or electronic form (e.g., paper chart or electronic data file), and since the health care provider owns the media in which the information is recorded and stored, the health care provider gains the property right of possession of data. \textbf{In essence, the health care provider becomes the legal custodian of patient's health care record and is given specific legal rights and duties relating to possession and protection of that health record.}

For the second question of ``who can read/write a patient's medical record?", we discuss the read and write right separately. For the write right of a medical record, we stipulate that only the creator of the medical record, who provides healthcare treatment or laboratory test for the patient and is responsible for the content of the medical record, can write data on it. \textbf{Once the data is written onto the medical record, it cannot be deleted and modified even when something goes wrong.} The healthcare provider can only create a new medical record to record the current diagnosis.

The read right of a medical record can be granted by the patient under the CDO's privacy policy. In general, patient himself and the creator of the medical record have the legal right to read the medical record. Other users, such as practitioners and insurance providers, should follow HIPAA's minimum disclosure principle. Depending on the identity or attributes of the user accessing the medical record, it should be further restricted which parts of the medical record he or she can access. For example, for an insurance provider, the portion that he can access may be limited to medical records to help reimburse medical expenses.

For the third question, when a patient is provided healthcare treatment in a CDO, he or she should sign a consent form in advance, agreeing that the owner of his or her medical records (CDO) grants other doctors the right to read his or her medical records under certain special circumstances. In addition, the patient also can sign an agreement that agrees that his or her medical records can be shared in a medical union to avoid repeated tests and unnecessary expenses. The patient or his loved one should be notified when the patient's medical record access rights are modified or granted to others.

For the last question, it mainly depends on how the CDO stores the large number of medical records. Most hospitals now have their own EHR system to store and manage medical records. Some EHR systems are built and operated by the CDO themselves and some EHR systems are built and operated in the cloud or a third party by purchasing storage service provided by a service provider. The former, which can be seen as a local storage method, has the advantage of being easy to manage, but it is difficult to achieve data sharing between EHR systems. The latter, which is a centralized storage, has the advantage that the CDO does not need to maintain the operation of the EHR system, but it loses control of the data because the data is stored in the cloud or a third-party database. Moreover, both these two methods are vulnerable to large-scale network attacks such as distributed denial of service (DDoS) attacks.

The blockchain technology developed in recent years provides possible solutions for secure distributed storage and sharing of electronic medical records. By combining distributed and cryptographic techniques, blockchain provides essential security attributes, such as authenticity verification, tamper resistance, and resistance to DDoS attacks, for secure data storage and data sharing.

\section{Blockchain: Overview}
\label{sec:bc}
The concept of blockchain was proposed by Satoshi Nakamoto in 2008~\cite{Nakamoto:2008:Bitcoin}. In 2009, Satoshi Nakamoto unveiled the first open-source implementation of blockchain - Bitcoin, which is a decentralized cryptocurrency system. Here, we briefly review the security and privacy in Bitcoin's blockchain lifecycle including its basic security and privacy properties and the techniques used to implement them.

\subsection{Consistency}
\label{subsec:consistency}
The concept of consistency in blockchain means that most nodes in the network have the same ledger during the given time period.
In Bitcoin, when a transaction is broadcasted by a node to the entire network, the miners will mine the transaction by packing it into a block along with other newly generated transactions and searching a nonce that satisfies the specific hash function (a.k.a. a proof of work challenge). Once a miner has completed its proof of work challenge, it sends its block and its proof to the network to solicit acceptances from other nodes. Other nodes will verify whether the nonce satisfies the hash function and all the signatures of transactions in the block. The other nodes accept the block by appending the block to its own replica of blockchain and using the hash of the accepted block as the hash of the previous block to generate the next block.

\textbf{Consensus.}
In order to ensure consistency and correctness of blockchain on a global scale, a scalable and secure consensus algorithm is required, which must ensure that (i) all copies of the blockchain maintain the same state at the same time, and (ii) the network offers a certain degree of fault tolerance and can prevent adversarial nodes from disrupting the consensus process without using a central authority.
Through a consensus protocol, transactions generated between any two nodes should be committed by most participants in the network. Also, the network should be resilient to the partial failures and ``attacks", such as when a group of nodes are malicious or corrupted.

\textbf{Gossip protocol} is also called epidemic protocol~\cite{1987:Epidemic}, which is mainly used for synchronizing data of each node in a distributed database system. It is an eventually consistent protocol. In the Gossip protocol, a certain node sends specified data to a group of other nodes in the network. The data spreads one by one like a virus through the nodes. Finally, the data is propagated to every node in the system. As a result, reliable data dissemination is realized in a large-scale distributed system. Bitcoin leverages the Gossip protocol to realize message dissemination and to achieve the consistency of messages across the network.

\subsection{Tamper-Resistance}
\label{subsec:tamper-resistance}

Tamper-resistance in the context of blockchain means that malicious users or adversaries cannot succeed in meddling with the information on the blockchain either during the block generation process or after the block is accepted by the entire network.
To achieve the goal of tamper-resistance, Bitcoin blockchain uses a cryptography technology named hash chain, which combines two existing and widely used techniques: hash pointer and Merkle tree.

\textbf{Hash pointer} is a compressed value of the transaction data by a hash function, pointing to the storage address of the data. By employing a collision-resistant hash function, a hash pointer has a capable of detecting whether the stored data has been modified. Blockchain uses hash pointers to link data blocks together. Specifically, each block has a hash pointer to the previous block, which stores the hash value of the previous block and points to the storage address of the previous block. In addition, users can publicly verify whether the transaction data in the storage has been tampered with by comparing the hash of the stored data with the hash of the original data. If the two hashes are inconsistent, then the stored data has been tampered with. By linking the root hash of the genesis block with the successor blocks through hash pointer, it equips the entire chain with the ability to resist tampering effectively.

\textbf{Merkle tree} is a tree-shaped data structure in which each leaf node is tagged with a hash of the data block, and nodes other than the leaf nodes are tagged with a cryptographic hash of its child nodes. Merkle tree can effectively and safely verify the content of large data structures, which is a generalization form of hash chain. The Bitcoin's blockchain uses the Merkle tree to implement efficient and non-tamperable storage of transactions.

\subsection{Authenticity}
\label{subsec:authenticity}

The authenticity of a transaction in Bitcoin contains two security aspects. The first meaning is that the transaction is real, that is, indeed one party transfers a certain amount of money to the other party's account, and the transaction is issued by the payer. Second, it also means prevention of double-spending attacks, that is, the coins contained in the transaction are not spent more than once.

\textbf{Digital signature.}
Bitcoin combines digital signature (a.k.a. Elliptic Curve Digital Signature Algorithm (ECDSA)\cite{Johnson:2001:ECDSA}) that can be publicly verified and consensus algorithm to guarantee the authenticity of transactions. Combining consensus protocols with signature ensures that all transactions are included in the blockchain and can be publicly verified by all users, so it is easy to be discovered by users when the coins has been double-spent or the transaction is illegal.

\textbf{Simplified Payment Verification (SPV)} refers to verifying bitcoin payments without downloading and running a full chain, but only saves the block headers of the longest chain. It greatly reduces the amount of storage and calculation required by the Bitcoin wallet client.

\subsection{Pseudonymity}
\label{subsec:pseudoymity}
Pseudonymity is the near-anonymous state in which a user has an identifier that is not their real name: a pseudonym. Pseudonymity allows users to communicate with others in a generally anonymous way.
In a Bitcoin system, the address of a user account is a hash of their public key. Users use the hashes of their public keys as their pseudo-identities to interact with each other without revealing their real identity. Although the use of pseudonyms instead of real identities can protect the privacy of users, there is still a risk of exposing user identity information in the Bitcoin system.

\textbf{Pseudonymity by Public Keys as Pseudonyms.}
In a network environment, in general, the public key can be regarded as the identity of the user. Take the digital signature we mentioned earlier as an example, the user confirms the authenticity of the signature by verifying the signature using the signer's public key. That is because the public key is bound to the identity information of the user (such as name and email address) by issuing certificates with a certificate authority (CA). In addition, public keys are managed by a trusted third party, a.k.a. the public key infrastructure (PKI). One can obtain the public key of a user from PKI and trust that the public key does belong to the user. With PKI, the process of verifying the signature is automatically converted to the process of verifying the identity of the signer. Therefore, in scenarios where PKI and CA are used, the public key can be considered as an identity.

While Bitcoin is deployed in a decentralized network, so that CA and PKI cannot be used to register a user and manage public keys in Bitcoin's system. In Bitcoin system, users generate private and public key pairs by themselves. Since without CA the user's identity information cannot be bound to the public key, these public keys generated by users themselves can be viewed as the user's pseudonym and independent of their identity information.

\begin{figure}[t]
\centering
\includegraphics[width=3.5in]{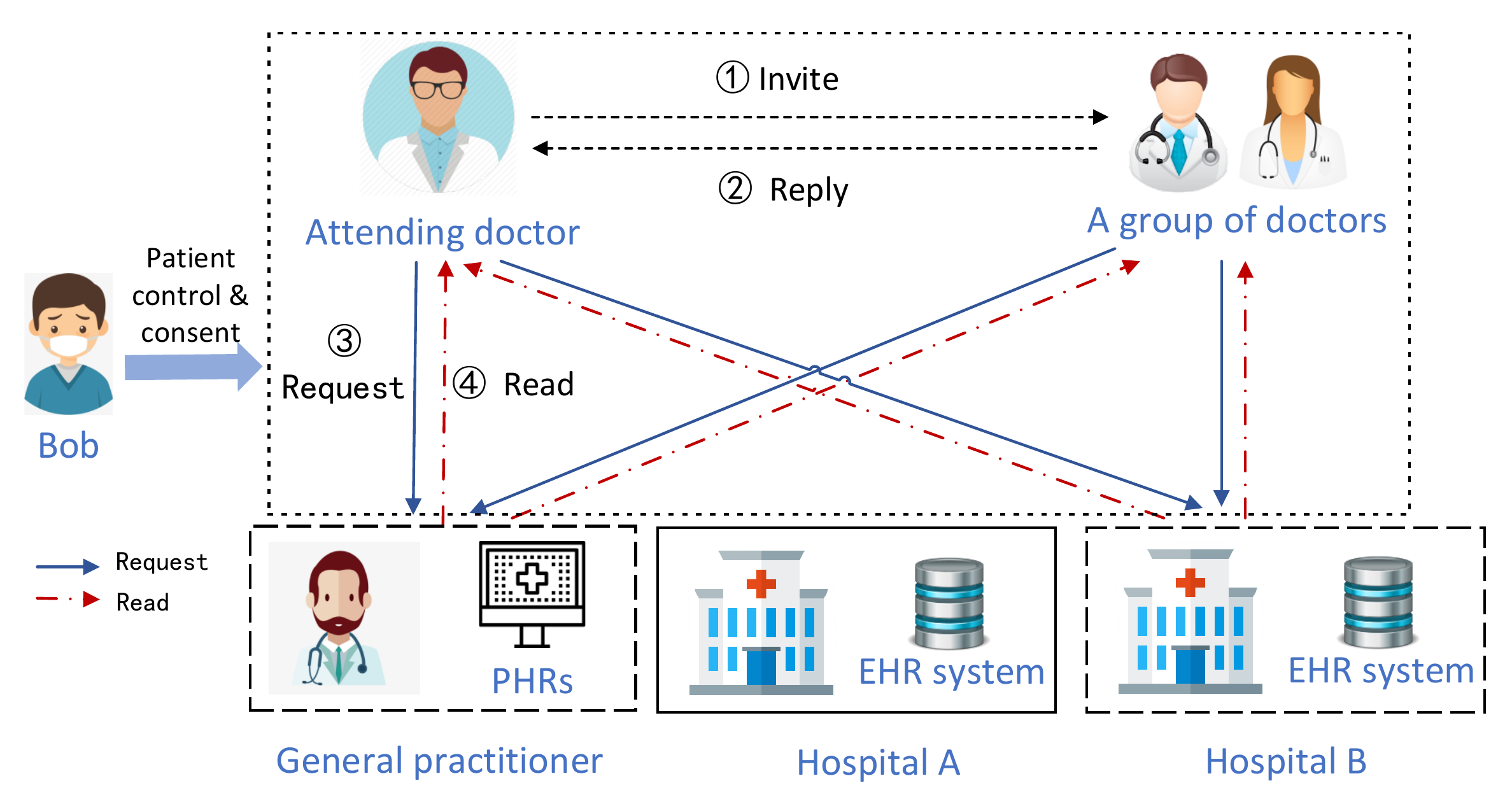}
\caption{An EHR Sharing Usage Scenario without Blockchain.}
\label{fig:usage}
\vspace{-10pt}
\end{figure}

\subsection{Unlinkability}
\label{subsec:unlinkability}
Unlinkability means that an adversary is unable to infer any relationship between two or more entities within the system after his observation with a-priori knowledge. Unlinkability is considered an anonymity property that provides more rigorous anonymity to the system when implemented in conjunction with pseudonymity.
In Bitcoin system, users can generate any number of random key pairs (multiple addresses) using a random algorithm, just as people can create multiple bank accounts at their own discretion. The difference is that these key pairs are generated by a random algorithm and are not bound to any identity information of the user, and there is no connection between them. Users use different pseudonyms to send and receive transactions, which achieve \textbf{partial unlinkability} of transactions.

Although Bitcoin uses multiple pseudonyms to make the transactions of same user seem unlinkable, it does not mean that Bitcoin can provide perfect anonymity. The user's transaction can still be connected to reveal the user's real identity information under the \textbf{de-anonymization inference attacks}~\cite{Narayanan:2016:BCT}.
Concretely, because each transaction is recorded with the sender's address and the recipient's address on the blockchain, anyone can use the sender's and recipient's address to track transactions freely by observing the total of bitcoins that are flowed into and out of the account. Thus, by mining the addresses in the transactions and the transaction flow of the bitcoins using statistical methods, the user's transactions can be linked to other transactions. More seriously, once a user's real identity is linked with her Bitcoin addresses, information of all transactions related to her bitcoin addresses will be revealed.

\section{Healthcare Blockchain: Overview}
\label{sec:health-bc}
In this section, we start with an EHR sharing usage scenario to illustrate that the blockchain can be used to solve the current cross-institutional sharing of EHRs. Then we discuss the basic concepts and taxonomy of healthcare blockchain.

\subsection{An EHR Sharing Usage Scenario}
\label{subsec:usage}
A patient, named Bob, felt that the stomach was uncomfortable and sometimes had a pain. His general practitioner, whom he regularly visits, suggested him to go to a hospital, say hospital A, for an in-depth examination. Unfortunately, he was finally diagnosed a gastric cancer. He was advised to transfer to a more specialized hospital (a cancer-treatment center, say hospital B) for surgical removal of the stomach (gastrectomy). After entering the hospital B, there is an attending doctor who is mainly responsible for the treatment of Bob. Because Bob also has diabetes, which causes some complications, his attending doctor seeks the advice from other experts from other hospitals to jointly develop Bob's surgical plan, including Bob's specific general practitioner, who is fully understands Bob's medical history.

\begin{figure}[t]
\centering
\includegraphics[width=3.5in]{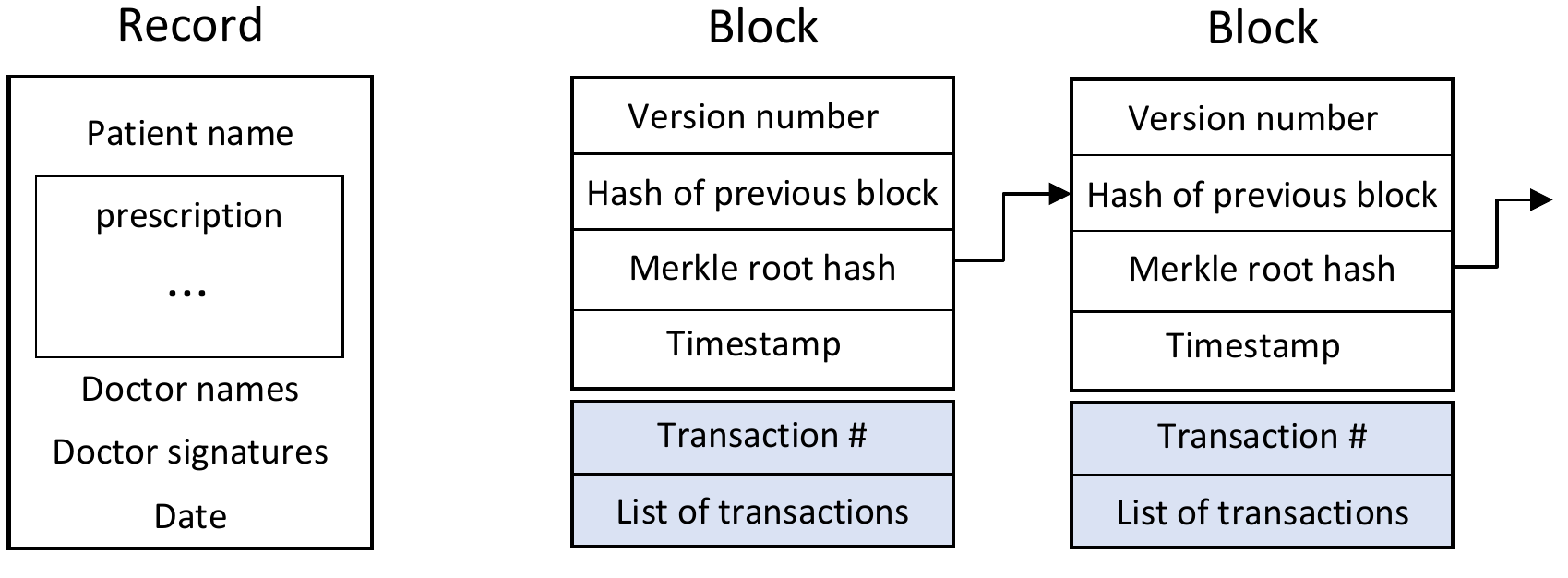}
\caption{The Structures of Record and Block.}
\label{fig:structures}
\vspace{-10pt}
\end{figure}

In this scenario, every participant of the consultation group needs to access Bob's medical records based on their areas of expertise and roles, thus all the access procedures should follow the Health Insurance Portability and Accountability Act (HIPAA) minimal disclosure principle. Moreover, because Bob suffers from other diseases, such as diabetes, in order to develop a more complete surgery and treatment plan, in addition to access the medical records and examination results generated in hospital B, participants of the consultation group should also be informed about Bob's medical history, which is recorded in his PHRs and EHRs owned by other CDOs.

\begin{figure*}[!t]
\centering
\includegraphics[width=7in]{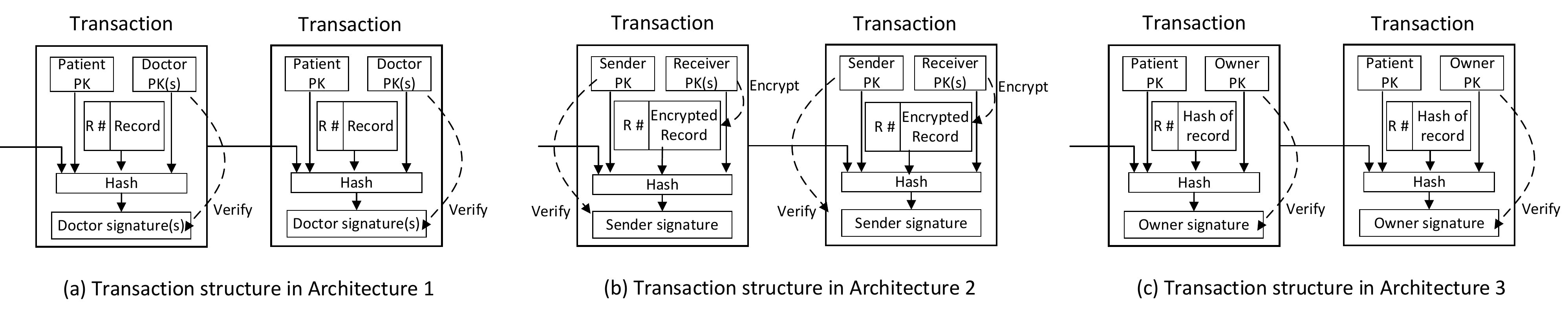}
\caption{The Transaction Structures of Three Reference Architectures}
\label{fig:transactions}
\vspace{-10pt}
\end{figure*}

For privacy protection and ease of management, a healthcare blockchain for efficient sharing of medical data can be used in this scenario. A legal user of the healthcare blockchain can read medical data stored on the blockchain. Specifically, the use of a blockchain to implement EHR sharing in this scenario has the following advantages:
(1) All EHRs of the patient are stored on the chain including Bob's historical medical records, each member of the consultation group can access easily and efficiently.
(2) All treatment-related actions, including EHR access logs, prescribing, etc., can be stored on the chain, so that patient and other legal users can access them at anytime and anywhere. In the event of a medical incident, it is also very easily to trace.
(3) In the healthcare blockchain, every EHR is signed by the doctor who created or is responsible for it and then hashed by a collision-resistant hash function. Finally, every EHR is packaged into a block and is appended to the chain by the techniques called Merkle tree and hash chain. By these techniques, it is impossibility to tamper with an EHR stored on the chain for a malicious user.

\subsection{Basic Concepts of Healthcare Blockchain}
\label{subsec:basic-concepts}
In this section, we clarify the conception of ``record", ``transaction" and ``block".

\textbf{Record} is a record of a patient's health and medical history. As shown in Fig.~\ref{fig:structures}, depending on the level or need of care a patient has, the type of records may vary, but all medical records will contain some common information such as patient's name, the doctor's names who create it and the generated date. In an electronic medical record (EMR), it is necessary to contain the digital signature of the doctor who is responsible for it. In the preceding usage scenario, after the group of doctors have discussed and determined the treatment plan, all of them or partial of them who are responsible for the content of record on behalf of all doctors in the consultation group are required to sign the record.

\textbf{Transaction} is a record that is broadcast to the blockchain network and collected into blocks. As shown in Fig.~\ref{fig:transactions}, the form of transactions can be different based on the different healthcare blockchain architectures. The detailed designing for each type of transaction will be discussed in Section~\ref{sec:architectures}. Unlike records, we replace the user's identity with a public key in a transaction. This partially protects the privacy of the user in the case of generating a public-private key pair using a random algorithm.

\textbf{Block} is a collection of transactions. Blocks are organized into a linear sequence over time. New transactions are constantly being processed by miners into new blocks which are added to the end of the chain. As shown in Fig.~\ref{fig:structures}, as well as a set of transactions, a block also contains hash of previous block, Merkle root hash, timestamp, the number of transactions, etc. Similarly, the structure of the block and the content it contains will vary depending on the used consensus algorithm.

In a healthcare blockchain system, we assume that a new record of patient is created by doctors in the EHR system owned by the CDO where the patient had treatment. When the patient agrees that the medical record can be shared with other medical institutions or individuals, the EHR system will transfer the record to transaction and broadcast it to the network. After the process of mining and consensus, a new block containing the transaction will be ``automatically" appended to the blockchain.

\subsection{Taxonomy of Healthcare Blockchain}
\label{subsec:taxonomy}
\begin{figure}[!t]
\centering
\includegraphics[width=3.4in]{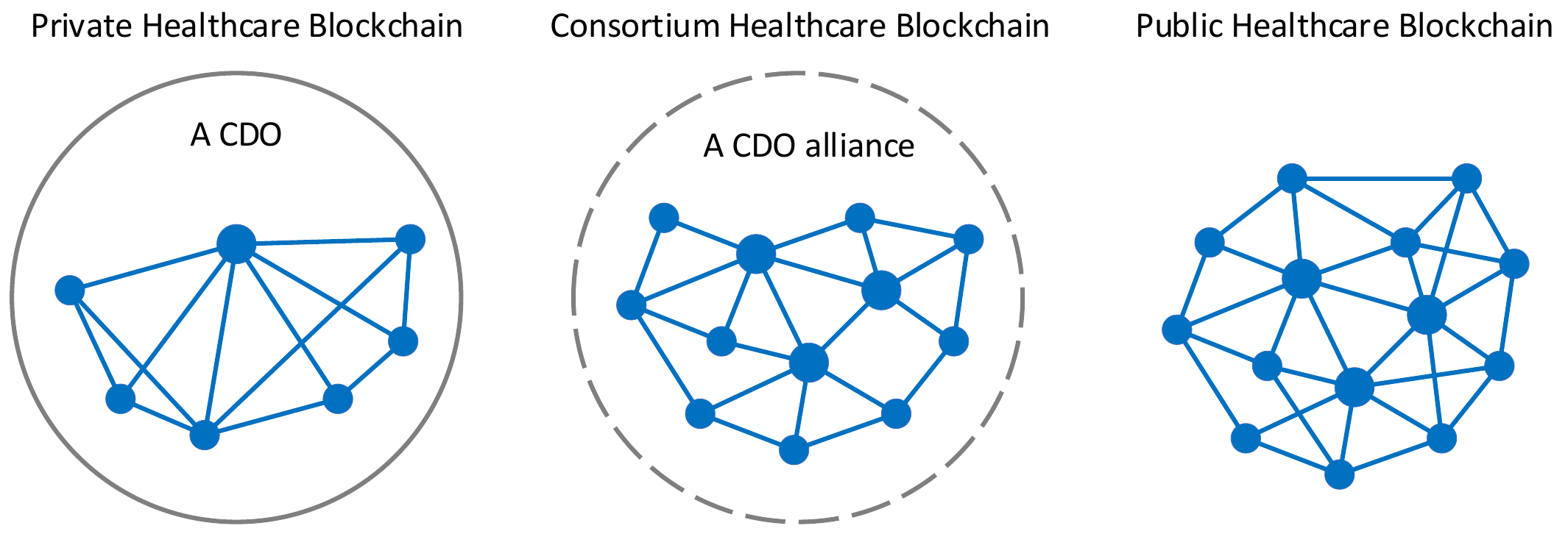}
\caption{Three Types of Healthcare Blockchain}
\label{fig:network}
\vspace{-9pt}
\end{figure}

According to the sharing range of data on the chain, the healthcare blockchain can be divided into three types as shown in Fig.~\ref{fig:network}.

\noindent \textbf{Private healthcare blockchain.}
The blockchain system is fully operated by a CDO, and the data stored on the blockchain is owned by the CDO. The blockchain infrastructure can be built and managed by CDO itself or provided by a third party who provides a blockchain platform.

\noindent \textbf{Consortium healthcare blockchain.}
The blockchain infrastructure and the data stored on it is shared by several CDOs and other relative organizations or individuals, such as healthcare provider organizations under a medical service group or within a region. Its infrastructure is likely to be built and managed by a trusted third party or blockchain platform provider.

\noindent \textbf{Public healthcare blockchain.}
The blockchain platform and the data stored on it is made available to the public. The blockchain infrastructure is built and owned by a blockchain service provider.

\begin{table*}%
\caption{Summarization of Representative Security and Privacy Requirements and Techniques of a Healthcare Blockchain}
\label{tab:sec-pri-requirement}
\begin{minipage}{\columnwidth}
\begin{footnotesize}
\begin{center}
\begin{tabular}{|c|p{3.5cm}|p{3.5cm}|p{3.4cm}|p{0.5cm}|p{4cm}|}
\toprule
 \multicolumn{2}{|c|}{\multirow{2}{4.6cm}{Security and privacy requirements}} &\multicolumn{2}{c|}{Supported by basic blockchain system (Bitcoin)} & \multicolumn{2}{c|}{Need to be enhanced}\\\cline{3-6}
 \multicolumn{2}{|c|}{}&\makecell[c]{Y/N} & \makecell[c]{Techniques} & Y/N & \makecell[c]{Potential techniques}\\\hline
 \multirow{5}{0.3cm}{\rotatebox{90}{Security}}&\makecell[c]{Consistency of transactions}  &\makecell[c]{Yes} &\makecell[c]{Consensus algorithms \\ Gossip protocol} &&\\
 &\makecell[c]{Integrity of transactions} &\makecell[c]{Yes} &\makecell[c]{Hash chained storage \\ Merkle tree} &&\\
 &\makecell[c]{Authenticity of transactions} &\makecell[c]{Yes} &\makecell[c]{SPV \\ Signature} && \\
 &\makecell[c]{Tracking and auditing} &\makecell[c]{Yes \\(access logs on chain)} &\makecell[c]{Hash chained storage\\ Signature verification} &&\\
 &\makecell[c]{Availability} &\makecell[c]{Partial \\ (resistance to DDoS attacks)} &\makecell[c]{Consensus algorithms} &\makecell[c]{Yes} &\makecell[l]{Security analysis and verification\\ of smart contracts}\\\hline
 \multirow{5}{0.3cm}{\rotatebox{90}{Privacy}}&\makecell[c]{Anonymity of users} &\makecell[c]{Partial \\(pseudonymity)} &\makecell[c]{Public key as pseudonyms} &\makecell[c]{Yes} &\makecell[c]{Anonymous signature}\\
 &\makecell[c]{Patient-control}&\makecell[c]{Partial \\(permission-on-chain)} &   &\makecell[c]{Yes} &\makecell[c]{Encryption \\(data encrypted by patient)}\\
 &\makecell[c]{Confidentiality} &\makecell[c]{No}  &   &\makecell[c]{Yes} &\makecell[c]{ABE, HE, SMPC, NIZK,\\ TEE-based solutions}\\
 &\makecell[c]{Fine-grained access control}&\makecell[c]{No}   &   &\makecell[c]{Yes}  &\makecell[c]{ABE, key management, HMAC}\\
 &\makecell[c]{Authentication of users} &\makecell[c]{No} & &\makecell[c]{Yes} &\makecell[c]{Anonymous authentication}\\
\bottomrule
\end{tabular}
\end{center}
\end{footnotesize}
\bigskip\centering
\end{minipage}
\vspace{-9mm}
\end{table*}%

Security and privacy are more than simply implementing user rights control and password authentication. Especially in the medical blockchain platform where patient sensitive information is stored, implementing multidimensional security and privacy attributes is a complex and challenging task. Security and privacy should be a top priority in the healthcare blockchain. We believe that patient data should be fully protected, including necessary cyber-attack protection, data encryption, user authentication and application security, as well as security operations and qualifications that meet the latest standards.

\section{Security and Privacy Requirements}
\label{sec:sp-HBC}
The data stored on the healthcare blockchain is more sensitive than Bitcoin's blockchain. People are more reluctant to disclose their medical information to others. Therefore, the developers and operators of healthcare blockchains should pay more attention to the protection of private information. We summarize security and privacy requirements of a healthcare blockchain as follows.

\noindent \textbf{Consistency of transactions.} The medical data in each transaction stored on node should keep consistent from each other. Obviously, this can be guaranteed by the consensus algorithm used in blockchain, which makes nodes confirm the new block and achieve consistency eventually by appending the new generated block to their own chain.

\noindent \textbf{Integrity of transactions.} The medical data packed into a transaction cannot be tampered with others during the process of broadcasting, mining, and storing on the blockchain. It can be achieved by the inherent hash chained storage mechanism of the blockchain. A collision-resistant hash function applied in the blockchain makes a malicious user almost impossible (with a negligible probability) to successfully tamper with transactions.

\noindent \textbf{Authenticity of transactions.} The users of a healthcare blockchain (readers of the medical record) need to confirm that the medical record in a transaction is authentic, that is, it is sent by a legal owner and has not been forged. In a transaction of a healthcare blockchain, it is required a digital signature signed by the owner of the medical data who generates and sends the transaction to blockchain network, so that users can confirm the authenticity of transaction by verifying the signature with sender's public key.

\noindent \textbf{Availability of system and transactions.} The availability of healthcare blockchain means that the users can access transaction data at anytime and anywhere, which contains system level and transaction level availability. The former refers to the fact that even in the case of a large-scale network attack, the system should run reliably. The latter means that the transactions stored on the blockchain is always accessible to users without being deleted, corrupted, and tampered with. For the former, the distributed network structure and consensus mechanism with Byzantine fault of the blockchain make the DDoS attacker must have sufficient computational resources to compromise most blockchain nodes. The larger the blockchain network, the more difficult it is for such a large-scale DDoS attack to succeed. For the latter, the structure of the hash chain of the blockchain ensures that transactions stored on it cannot be deleted and tampered with by malicious users.

\noindent \textbf{Anonymity of users.} Users may ask for anonymity in the process of authenticating and sending medical data. Because the data on the blockchain can be accessed by any legitimate user, the user does not want to associate the information of the medical data with his identity. More seriously, once one piece of medical information of a user is linked to his identity, it will lead to the leakage of all his medical data, thus affecting his life and work.

\noindent \textbf{Patient-control.} Patient-control includes (1) it should be controlled (or agreed) by the patient who can access which part of his or her medical data; (2) the system administrator or blockchain users cannot forward the medical data to anyone without the patient's permission.

\noindent \textbf{Confidentiality of transactions.} The confidentiality means (1) unauthorized user cannot successfully read or infer any private information from the data stored on the blockchain; (2) even in the event of an unexpected failure or malicious network attack, the confidentiality of the patient's medical data should always be guaranteed.

\noindent \textbf{Fine-grained access control of transactions.} When sharing medical data with others, patients may require the minimal disclosure of their medical records and related information in a healthcare blockchain system. The minimal disclosure involves three related aspects: (1) the medical data can only be accessed by authorized users; (2) authorized users can only access medical data that they need; (3) authorized users can only access medical data of a patient during the granted legal (treatment) periods. Consider the preceding EHR sharing scenario as an example, the group of doctors can only access Bob's medical data they need after being authorized and during the consultation and treatment of Bob. When the treatment is over, they will no longer be able to access Bob's medical data.

\noindent \textbf{Tracking and auditing of access.} In an online medical data sharing system, once patient's medical record has been accessed, the patient should be notified immediately to let the patient know when his medical record was accessed by whom. In addition, the patient should be able to view the access log at any time and check if his or her medical data is properly accessed and used according to the privacy policy.

\noindent \textbf{Authentication of users.} Usually, the healthcare blockchain is deployed in the form of permissioned blockchain, which has a management node to authenticate users. Thus, in a permissioned healthcare blockchain, on the one hand, it is required to effectively authenticate users to prevent illegal users from accessing medical data, and on the other hand, users do not want to reveal their real identity to the management node in the process of authentication.

We summarize the security and privacy requirements and corresponding techniques in Table 1. We observe that some requirements are guaranteed by the inherent properties of blockchain, such as consistency, tamper-resistance, and authenticity. Other security and privacy requirements can be partially achieved by using blockchain, such as availability and anonymity, and need to be enhanced by incorporating additional technologies into the blockchain. The support of the remaining security and privacy requirements need to be fully implemented using other technologies.

\begin{figure*}[!t]
\centering
\includegraphics[width=7in]{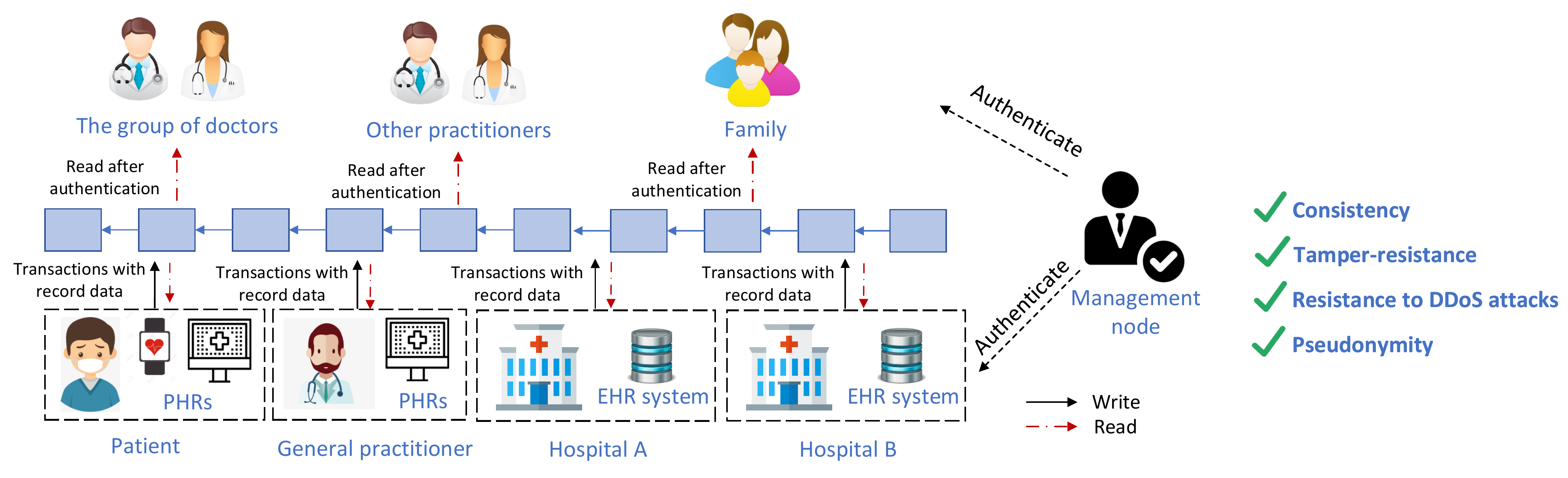}
\caption{Scenario 1: Data Storage and Sharing Chain with Plaintext-on-Chain}
\label{fig:arch-1}
\vspace{-10pt}
\end{figure*}

When we discuss the security and privacy requirements of healthcare blockchain, we need to choose different level of security and privacy requirements based on different application contexts of the healthcare blockchain. Especially in terms of anonymity, the choice of the degree of anonymity is related to the sensitivity of the data stored in the blockchain, the patient's requirements and other factors.

\noindent \textbf{Identifiability vs. anonymity.} In a healthcare blockchain system, on one hand, patients wish to keep completely anonymous when a receiver (or accessor), such as a researcher, accesses and uses part of the patients' medical information without knowing the patients' real identity. On the other hand, when an accessor is a doctor who want to search the medical records of his or her patient, he or she needs to know the identity of the sender in order to accurately find the patient's medical records.

\noindent \textbf{Linkability vs. unlinkability.} Similarly, in some cases, patients do not want their medical records to be linked. For example, a researcher who want to access all medical data about heart disease to see the relationship between heart disease and age, gender, and region. According to the minimal disclosure policy, the researcher should not access any other disease information by linking patients' medical records together. While, in other cases, patient want their medical records to be linked to form a chronological medical record chain at all CDOs across inpatient and outpatient. This makes the patient self and doctors who provide treatment to the patient easily to find a certain medical record.

Here raises a question: a publicly accessible healthcare blockchain often contains users with multiple roles, so what kind of anonymous property should such a healthcare blockchain provide? Given the particular nature of anonymous requirements for healthcare blockchains, we suggest that users of a healthcare blockchain use a keyed-hash message authentication code (or hash-based message authentication code (HMAC))~\cite{Bellare:1996:KHF} to generate pseudonyms. A user may have several pseudonyms with different keys. When a user wants someone to link his or her medical records, he or she can release the keys to the person. When he does not want the user to access his future medical records, he just needs to generate a new pseudonym with a new key. The more frequently a user replaces a key, the less medical records can be linked. A more reasonable approach is to use one pseudonym when Bob is being provided treatment in hospital A and release the key which used to generate current pseudonym to the doctors of hospital A. Bob replaces his pseudonym with a new one when he has transferred to hospital B. By this method, doctors of hospital A can only access the medical records created in hospital A and cannot access the medical records created in hospital B.

\section{Scenarios for EHR Sharing}
\label{sec:architectures}
In this section, we categorize existing efforts into three reference blockchain usage scenarios for EHR sharing based on different forms of EHR data stored on the chain and describe each of them in detail.

\subsection{Data Storage and Sharing Chain}
\label{subsec:scenario-1}
The first scenario is shown as in Fig.~\ref{fig:arch-1}. In this scenario, transactions with plaintext of medical record are directly stored on the blockchain. Therefore, all users of the blockchain system can access all the medical records stored on the blockchain. Since the blockchain stores plaintext medical records, it needs to have a trusted third party responsible for authenticating and revoking users. Individuals such as patients and doctors can be granted access right once they have been authenticated. This scenario is suitable for the private healthcare blockchain or consortium healthcare blockchain. The administrator of the blockchain or a trusted third party can act as a management node for authenticating and revoking users. We name this scenario data storage and sharing chain (DSS chain for short).

Take the preceding EHR usage scenario as an example, both hospital A and hospital B are in a medical organization alliance, and the medical organizations that join it can share medical data with each other by a shared healthcare blockchain system. The group of doctors as well as the patient Bob and his family can access the medical records stored on the blockchain as long as they have been authenticated by the management node.

The reference transaction structure of DSS chain is shown as in Fig.\ref{fig:transactions}(a).
It contains a patient's public key, the set of doctors' public keys who create the record, and the content of record. In addition, a collision resistant hash with the patient's previous transaction, the patient's public key, the content of record and the public keys of doctors as its inputs is needed to achieve tamper-resistant of transactions. This forms a chronological medical record chain of a patient at all CDOs belonging to the alliance across inpatient and outpatient. The output of the hash function is signed by doctors who are responsible for the content of the record with their private keys.
Users can verify the signatures with the public keys contained in the transaction.

\subsubsection{Properties of DSS Chain}
\label{subsubsec:pro-DSS}
The advantages of DSS chain can be summarized as follows: (1) The medical records stored on the blockchain can be efficiently shared with users in the system. (2) Each transaction is linked to the previous transaction of the patient. It forms a chronological medical record chain of a patient at all CDOs belonging to the alliance across inpatient and outpatient. This not only makes doctors conveniently to track the patient's condition, but also supports users easily to find a certain record.

While, since the medical records are stored on the blockchain in a plaintext form, all users of system can access all the records stored on it. If
there are malicious users in the system, it will lead to information leakage. Moreover, the process of authentication also will leak privacy information to the administrator of the system. Therefore, additional security and privacy techniques such as symmetric encryption and anonymous authentication are needed to enhance the security and privacy of such healthcare blockchain.

\subsubsection{Enhanced Techniques of DSS Chain}
\label{subsubsec:tech-DSS}

\noindent \textbf{Anonymous Authentication:}
Anonymous authentication is a technology that enables users to be authenticated under the condition of privacy-preserving. Researches on anonymous authentication schemes can be classified into: three-party schemes~\cite{Zhu:2004:ANA,Gope:2016:LEE} and two-party schemes~\cite{Ni:2016:AMA,Yang:2010:UAP}.
In~\cite{Gope:2016:LEE}, an efficient anonymity authentication scheme was proposed by Gope et al. for roaming services in global mobility networking. Their scheme is full-filled strong user anonymity and greatly secure and efficient due to the hash function and bitwise exclusive-or operation was employed. Yang et al. introduced the concept of two-party anonymous authentication scheme and proposed two schemes for roaming services in~\cite{Yang:2010:UAP}, which are constructed by identity-based signature and group signature, respectively. Their schemes can efficiently achieve user anonymity and user revocation.
In~\cite{Yang:2019:AnFRA}, an anonymous and fast roaming authentication scheme for space information network (SIN) was proposed. Their scheme is constructed by group signature, which not only makes users can be authenticated without the participation of the home server, but also provides strong users anonymity and guarantees its security requirements. In~\cite{Djellalbia:2016:AAS}, an anonymous authentication scheme in the cloud environment for e-health has been proposed.

\begin{figure*}[!t]
\centering
\includegraphics[width=6.3in]{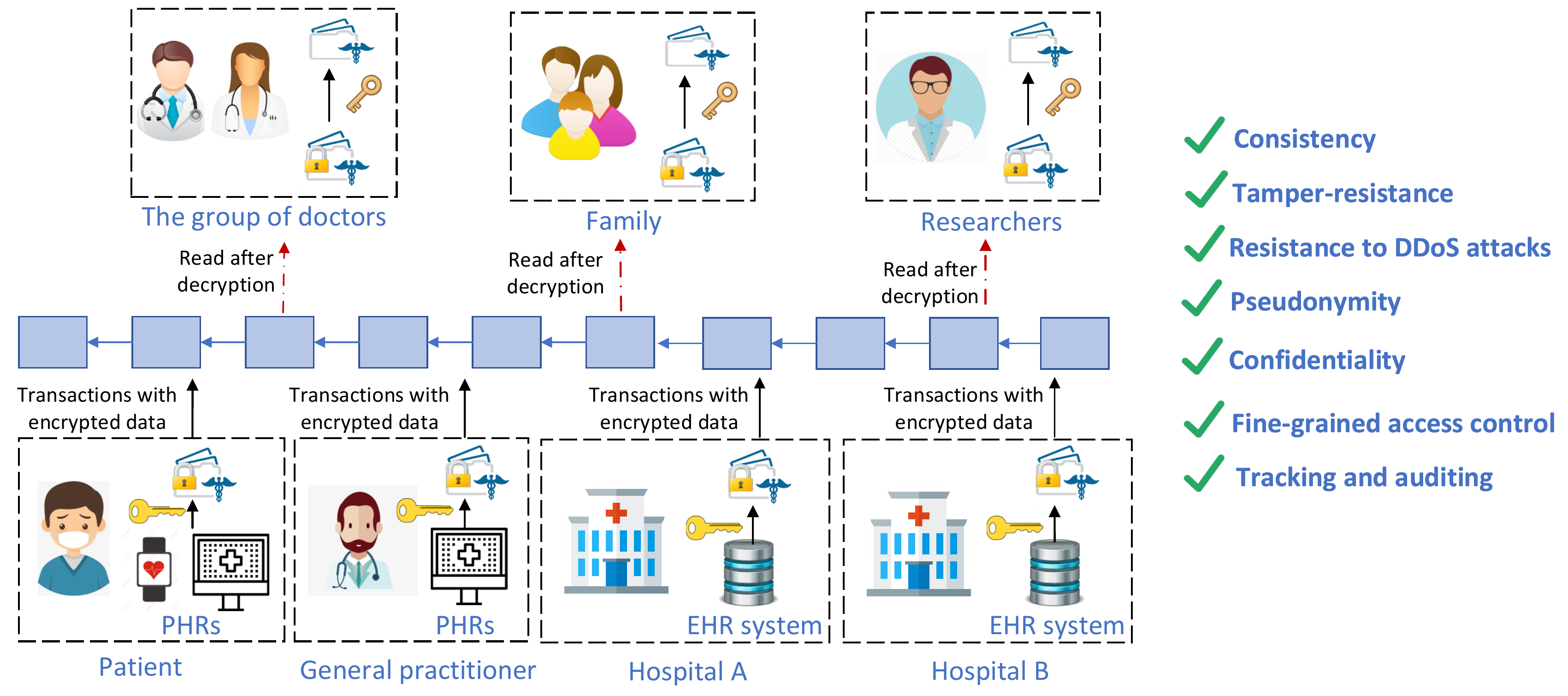}
\caption{Scenario 2: Secure Transmission and Tracking Chain with Ciphertext-on-Chain}
\label{fig:arch-2}
\vspace{-10pt}
\end{figure*}

\noindent \textbf{Anonymous Signature:}
The anonymous signature algorithm refers to a signature algorithm that can provide anonymity to the signer. Group signature, ring signature and attribute-based signature are the most typical anonymous signature algorithms.

\noindent \textbf{(1) Group Signature.}
Group signature is proposed by Chaum and Heyst in 1991~\cite{Chaum:1991:GS}. Given a group, any of its members can sign a message for the entire group anonymously by using her own secret key, and anyone with the public key of the group can verify the generated signature and confirm that the signature of some group member is used to sign the message. The signature verification process cannot reveal the signer's true identity, except that the signer is a member of the group. Therefore, for a well-designed group signature algorithm, assuming that there are $n$ members in the group, the probability of the opponent successfully guessing the signer is $1/n$. Obviously, the more group members, the less probability the adversary will succeed in guessing the signer.
The group signature algorithm has a group manager who is responsible for managing group members, such as adding and revoking group members, and publicizing the signer's identity in the event of a dispute.
Therefore, group signature can be applied in the consortium blockchain, the management node can be act as the group manager to manage the group.

\noindent \textbf{(2) Ring Signature.}
A ring signature~\cite{Rivest:2001:HLS} is another signature algorithm that can hide signer information into a group of users.
This signature algorithm uses a ring structure. Similarly, for a well-designed ring signature algorithm, the probability of the adversary successfully guessing the signer is $1/n$, where $n$ is the size of the ring, i.e., the number of ring members.
Unlike group signatures, ring signatures do not have a manager to set up and revoke the group and members and reveal the signer's identity in the event of dispute. A user who wants to sign a message with ring signature can organize a ``ring" by his own choice without additional communication with others.
Thus, ring signature algorithm is more suitable for public blockchain, which has no central authority party.
CryptoNote~\cite{CryptoNoteSig:2012} is one of the typical applications of ring signatures in blockchains. It uses the ring signature to hide the sender's public key by using the public keys of other users in the ring, so that it is difficult to identify the signer when verifying the signature. In 2015, Ethereum~\cite{Ethereum} provided ring signature option in its platform to ensure anonymity to users.

\noindent \textbf{(3) Attribute-Based Signature.}
Attribute-based signature (ABS) is a signature algorithm that allows users to sign messages with any predicate of their attributes issued from attribute authorities. ABS translates the verification of signer's identity into the verification of signer's attributes, thereby protecting the signer's true identity. In ABS, users cannot forge signatures with attributes they do not possess even through colluding.
The ABS can validate the attributes of the signer while concealing the true identity of the signer.
However, in practical applications, a user may have multiple attributes, and the certificates of these attributes may be issued by different administrative organizations. For example, a signer might sign an EHR with the private keys related to attributes ``doctor" and ``researcher", where the certificate of attribute ``doctor" is released by a healthcare organization and the certificate of attribute ``researcher" is emitted by a research institution. The verifier can verify that the signer has both attributes by verifying the signature. In the above application scenario, a trusted third party that can cross different organizations is required to authenticate the user's attributes and issue private keys.
Recently, inspired by the distributed attribute-based encryption (DABE) scheme proposed by Allison and Brent~\cite{Lewko:2011:DAE}, You Sun et al.~\cite{Sun:2018:DABS} proposed a decentralized ABS scheme for securing the healthcare blockchain. In their scheme, the attribute certificates and the corresponding signature keys are issued by each attribute authority, so that a third trusted party is not required as the central authority.

\subsection{Secure Transmission and Tracking Chain}
\label{subsec:scenario-2}
The second scenario is shown as in Fig.~\ref{fig:arch-2}. In this scenario, transactions with ciphertext of medical record are stored on the blockchain. Specifically, the record is encrypted by a secure asymmetric encryption algorithm with the receiver's public key. Then the ciphertext as well as sender's public key and receiver's public key are packed into a transaction and broadcasted to the network.
Only the user who has the corresponding private key can decrypt the record. Therefore, this type of healthcare blockchain system can support fine-grained access control without a trusted third party acting as a management node and is more suitable for consortium healthcare blockchain and public healthcare blockchain. Thus, in this scenario, the blockchain is mainly used as a medium for data security transmission. We name it secure transmission and tracking chain (STT chain for short).

Take the preceding EHR usage scenario as an example, when the group of doctors need to look at Bob's historical medical records to determine if he can be performed the surgery, Bob's general practitioner respectively encrypts each record by an asymmetric encryption algorithm with public key of each doctor in the group and broadcasts the generated transactions to the network. The group of doctors can decrypt each record with their private keys. In this scenario, an attribute-based encryption can be used to encrypt the patient's record, such that the sender only needs to encrypt once. This can significantly reduce the storage space for the blockchain. Similarly, when the group of doctors generate a record for Bob, they can encrypt the record with Bob's public key, create a signature with their private keys and send the transaction to Bob by the blockchain.

\begin{figure}[!t]
\centering
\includegraphics[width=3.4in]{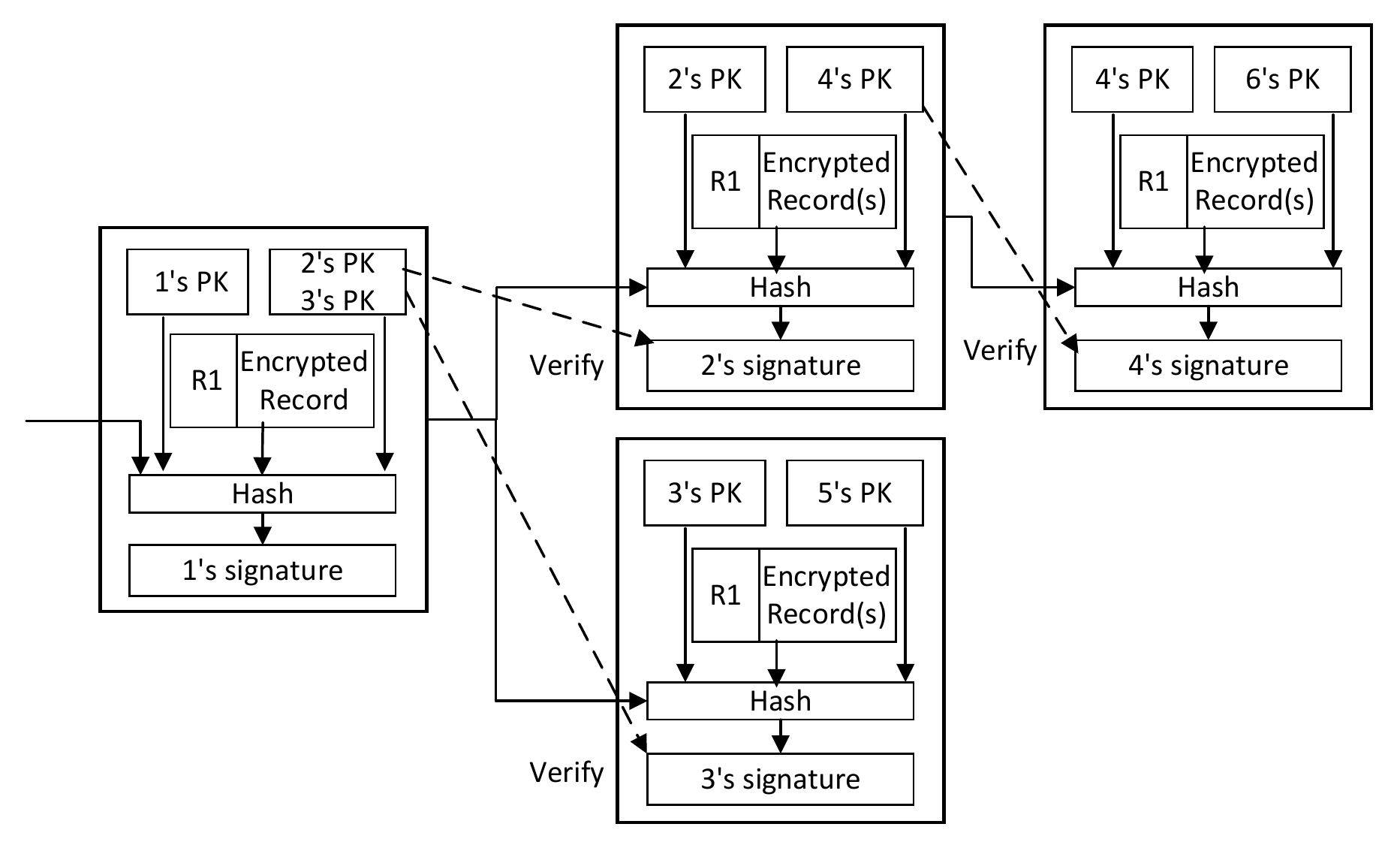}
\caption{An Example of Transmission Directed Graph}
\label{fig:trans-chain}
\vspace{-10pt}
\end{figure}

The reference transaction structure of STT chain is shown as in Fig.\ref{fig:transactions}(b).
It contains a sender's public key, receivers' public keys, the record number, and the ciphertext of the record encrypted with the receivers' public keys. Each sender transfers the record to the transaction by digitally signing a hash of the previous transaction and above values. A receiver can verify the signatures to verify the chain of transmission. Since we allow multiple recipients in this transaction structure, it actually forms a transfer directed graph of each record. An example of transmission directed graph is as shown in Fig.~\ref{fig:trans-chain}. A transaction containing record $R1$ is encrypted and sent to User 2 and User 3 by User 1, then User 2 and User 3 respectively sent $R1$ to User 4 and User 5. Other users in the system can verify the signature with the public key in its previous transaction. This forms a transmission directed graph of record $R1$, which can be used to track the transfer process of the record for patients.

\subsubsection{Properties of STT Chain}
\label{subsubsec:pro-STT}

The advantages of STT chain are as follows: (1) The medical records stored on the blockchain are encrypted, only the users who have corresponding private keys can decrypt and read the record. Thus, it can support fine-grained access control and can be deployed in the public healthcare blockchain. (2) Each transaction is linked to the previous transaction of transmission. It forms a chronological transmission directed graph for each record. This makes patients easily to track the transmission process of his own medical records. In STT chain, attribute-based encryption (ABE) can be applied in this architecture to achieve one-to-many transmission and sharing; Proxy re-encryption (PRE) can be used to achieve the transfer of read right without re-encrypting by the sender; Homomorphic encryption (HE) can be used to implement the operation on ciphertexts without revealing the plaintexts used for calculation.

While, since the medical records are stored on the blockchain in a ciphertext form, when the sender wants to share the medical data with other users he needs to re-encrypt the data, re-pack them into a transaction and broadcast to the network. It means that every transfer of a record will generate a transaction, which increases the storage complexity of the blockchain. In addition, collusion attack may be initiated by two or more colluding users by giving the key of the user with access right to another user. Therefore, additional security and privacy techniques such as ABE that is secure against collusion attack and PRE are needed to enhance the security and privacy of such healthcare blockchain system.

\subsubsection{Enhanced Techniques of STT Chain}
\label{subsubsec:tech-STT}

\noindent \textbf{Attribute Based Encryption:}
ABE, which was first proposed by Sahai and Waters ~\cite{Sahai:2005:FIBE}, is a special form of identity-based encryption that uses public keys related to attributes instead of a public key on the identity to generate a ciphertext. The encrypted message can only be decrypted if the decrypter has private keys corresponding to the required attributes. A secure ABE scheme should be against collusion attacks, that is, even if a malicious user collaborates with other users to obtain private keys that are not related to his attributes, he cannot successfully decrypt a message that the required attributes do not match his attributes.

In order to implement the issuance of attribute certificates for users' multiple attributes in a fully distributed network environment, a decentralized ABE scheme was proposed~\cite{Lewko:2011:DAE} in 2011, which is called DABE.
In \cite{Lewko:2011:DAE}, the key pairs on the attributes are issued by multiple attribute authorities and does not require the participation of a fixed central attribute authority. For instance, a patient wants only those users who are both doctor and researcher to access his EHR, so he uses the public keys corresponding to attribute predicate ``doctor AND researcher" to encrypt his EHR. Therefore, the EHR of the patient can be decrypted only when the user has both credentials: (1) the private key for the attribute ``doctor" and the private key for the attribute ``researcher". The attribute certificate and key pair associated with the ``doctor" are issued by a CDO, and the attribute certificate and key pair associated with the ``researcher" are issued by a research institution. While DABE can be well applied to the above scenario, so that keys issued by different attribute authorities can be combined to encrypt and decrypt messages. DABE offers the possibility for attribute encryption algorithms to be applied in blockchains.

\noindent \textbf{Homomorphic Encryption in Blockchain:}
The homomorphic encryption~\cite{Gentry:2009:FHE} is a powerful public key encryption technology. The term ``homomorphic" means that the encryption enables certain types of calculation to be performed directly on the ciphertext, and the result of the decrypting the computed value is the same as the result of performing the same calculation on the plaintext.
Using homomorphic encryption technology, users can directly store ciphertext on the blockchain and perform some calculations without any changes to the blockchain itself, thus providing users with the confidentiality and privacy protection of the data. Ethereum Smart Contracts provides homomorphic encryption for users that enables privacy calculations of encrypted data stored on it for better security and privacy.

Homomorphic encryption can be applied in healthcare scenarios. For example, the hospital encrypts the medical expenses of the patient and stores them in the blockchain, and the insurance company directly calculates the total payment cost on the encrypted data without revealing the details of the patient's expenses. Another typical application of homomorphic encryption in healthcare sector is privacy-preserving federated learning for healthcare research, which we will introduce in Section~\ref{subsec:blockchain-DL}.

\noindent \textbf{Proxy Re-encryption:}
In the preceding EHR usage scenario, when the patient Bob loses consciousness, how a doctor in the consultation group accesses the medical records encrypted with Bob's public key?
A proxy re-encryption can be used in the above special scenario to solve the problem of re-authorization of access rights in the case of patient unconsciousness. Bob could designate a proxy, say his general practitioner or his family members, to re-encrypt his medical records that is encrypted with Bob's public key without revealing Bob's private key to the group of doctors. In the process of re-encryption, the proxy generates a new key that a doctor in the consultation group can use to decrypt the medical records without revealing the contents of Bob's medical records to the proxy.

PRE was first proposed by Blaze, Bleumer, and Strauss in 1998~\cite{Blaze:1998:PRE}, which allows a semi-trusted proxy compute a function that converts ciphertexts for one party into ciphertexts for another party without seeing the underlying plaintext.
Very recently, some researchers and companies combine PRE with blockchain to implement access control and rights management.
The NuCypher network uses the Umbral threshold proxy re-encryption scheme to provide cryptographic access controls for distributed apps and protocols~\cite{Egorov:2018:NuCypher}. Manzoor et al.~\cite{Manzoor:2018:BBP} proposed a blockchain based proxy re-encryption scheme, which uses an efficient PRE scheme to control the sharing of IoT data. Chen et al.~\cite{Chen:2018:TPRE} presented a consortium blockchain access permission scheme based on the threshold PRE scheme.

\noindent \textbf{Non-Interactive Zero-Knowledge Proof:}
Zero-knowledge proofs proposed in 1985~\cite{Goldwasser:1985:KCI} is a privacy-preserving cryptographic technology. The basic idea is that it can fully prove that an assertion that certifier has stated is correct without leaking the relevant information, that is, for the verifier, the ``knowledge" of the assertion is ``zero".
The non-interactive variant of zero-knowledge proofs (abbreviated as NIZK) first proposed by Blum, Fledman and Micali~\cite{Blum:1988:NZA} is a non-interactive form of zero-knowledge proofs.
By sharing a common reference string (CRS) between the certifier and the verifier, NIZK can implement zero-knowledge proof without requiring real-time communication between the certifier and the verifier.
When the zero-knowledge proof is applied to the blockchain to carry out the transaction, it allows the sender of the transaction to prove to the recipient that the sender has enough balance to pay for the transaction through zero-knowledge proof without having to disclose his account balance.

In 2012, Bitansky et al~\cite{Bitansky:2012:ECR} proposed a more efficient NIZK protocol, a.k.a. zero-knowledge Succinct Non-interactive ARgument of Knowledge (abbreviated as zk-SNARK). The proofs of zk-SNARK are particularly short and easy to verify. It has been adopted by Zcash~\cite{Sasson:2014:ZDA} as a cryptographic engine to protect the privacy of users.
Recently, the Ethereum R\&D team and the Zcash team has collaborated to address the combination of programmability and privacy in the blockchain. The key to adding such capabilities to Ethereum is to enable using a variety of zk-SNARK constructions on Ethereum blockchain.
However, when using SNARK, a complex setup phase is required to build system common parameters. These common parameters need to be generated in a secure manner every time to use a SNARK for a particular circuit. This greatly hinders the availability of SNARK. Simplifying this setup phase is an important challenge for SNARK to be widely deployed on blockchain applications.

\begin{figure*}[!t]
\centering
\includegraphics[width=6.5in]{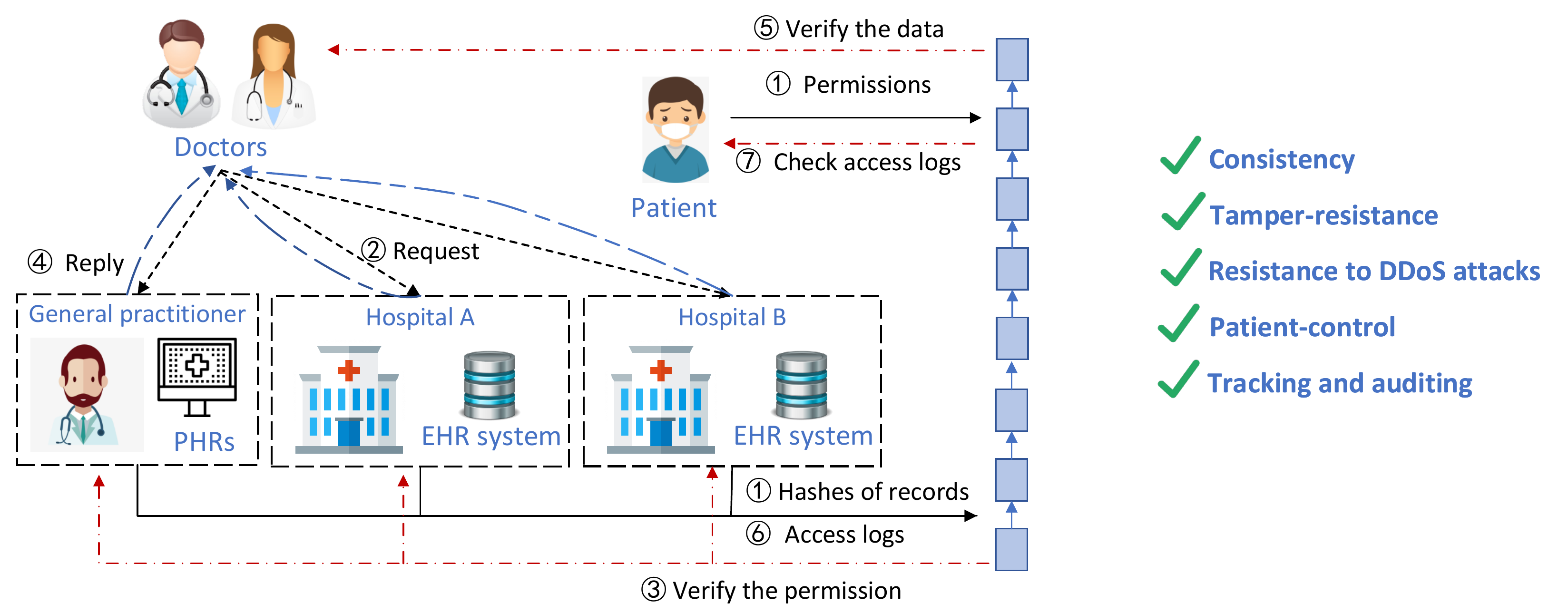}
\caption{Scenario 3: Storage and Permission Verification Chain with Hash-on-Chain}
\label{fig:arch-3}
\vspace{-10pt}
\end{figure*}

\noindent \textbf{Secure Multi-Party Computation (MPC):}
MPC allows multiple parties to participate in the computation without revealing the inputs of each party. This ensures that the adversary can only observe the result of the computation output and cannot obtain the inputs of computation participants.
In 1982, Andrew Yao first proposed and formally defined secure two-party computation (2PC) protocol~\cite{Yao:1982:PSC}. In 1987, Goldreich and his coauthors extended the 2-party computation to the $n$-party computation ($n>2$)~\cite{Goldreich:1987:HPM}, which can be seen as the basis for many efficient MPC protocols.
As early as 2008, MPC was first deployed on a large scale online auction system to protect the bid privacy of bidders in Denmark~\cite{Bogetoft:2009:SMC}. Since then, MPC has been successfully applied to multiple real-world systems, such as distributed voting, private bidding, and private information retrieval to protect the privacy of system users.

In recent years, researchers and technicians are also exploring the application of MPC in blockchain systems. In 2014, the first secure multiparty computing protocols that could be applied to the Bitcoin system were designed and implemented by Andrychowicz and his colleagues~\cite{Andrychowicz:2014:SMC}. They designed decentralized secure multiparty protocols for lotteries without the involvement of any trusted third party during the operation of the protocols. Their protocols guarantee the fairness of honest users in the presence of dishonest users.
Another typical application of MPC in the blockchain is a decentralized SMP computing platform called Enigma, which is designed in 2015~\cite{Zyskind:2015:Enigma}. Enigma combines blockchain technology, secure multi-party computing technology and other security technologies to achieve high-level security and privacy. Concretely, in order to achieve privacy for its computation model, the SMPC is implemented with a verifiable secret sharing scheme.

\subsection{Storage and Permission Verification Chain}
\label{subsec:scenario-3}
The third scenario is shown as in Fig.~\ref{fig:arch-3}. In this scenario, transactions with hash of medical record are stored on the blockchain. Specifically, a medical record is generated and stored on the EHR system maintained by a CDO, and meanwhile a hash value of the new generated medical record is calculated and broadcasted to the blockchain network within a transaction.
Individuals such as patients and doctors request access to medical records from EHR systems. EHR system replies the access requests with medical data according to the pre-defined privacy-preserving policies. Users can verify whether the data obtained from the sender has been tampered with by comparing the hash value of received record with the hash value stored on the blockchain. In this healthcare blockchain architecture, only the hashes of medical records rather than the medical records are stored on the blockchain. Thus, this architecture realizes the public verification of data by using the tamper-resistant property of blockchain and achieves security storage of medical records. In addition, blockchain in this scenario can also be used to store access permissions predefined by patients and access logs so that patients can easily check who accessed their medical records and whether the CDOs have adhered to pre-defined privacy policies. Thus, we name it storage and permission verification chain (SPV chain for short). Obviously, this architecture can be deployed on the public blockchain.

Take the preceding EHR usage scenario as an example, when Bob visits a doctor in a CDO, the EHR system automatically broadcasts the corresponding hash of the medical record to the blockchain network as soon as a new medical record is generated by the doctor. Therefore, it forms a chronological chain of hashes of medical records for Bob. Meanwhile, Bob broadcasts access permissions of his medical records to blockchain network. When the group of doctors need to access the medical records stored on each EHR system, they respectively request to access medical data from each EHR system. EHR system first checks the access permission stored on the blockchain, and then replies them with medical data according to the pre-defined privacy-preserving policy. When the doctors in consultation group received medical records, they can check whether or not the records were tampered with by comparing the hashes of medical records with the hashes stored on the blockchain. In addition, hospital B stores the access logs of Bob's medical records on the blockchain. Bob can check if the hospital and the group of doctors have followed the privacy policy.

The reference transaction structure of SPV chain is shown as in Fig.\ref{fig:transactions}(c).
It contains a patient's public key, the owner's public keys, an unique number of the record and the hash of the record. In addition, a collision-resistant hash with the patient's previous transaction, the patient's public key, the number and the hash of the record and the owner's public key as its inputs is needed to achieve tamper-resistant of transactions. This forms a chronological chain of hashes of medical record for a patient at all CDOs. The output of the hash function is signed by owner who stores the record in its EHR system with its private key.
Users can verify the signatures with owner's public key.

\subsubsection{Properties of SPV Chain}
\label{subsubsec:pro-SPV}

The advantages of SPV chain can be summarized as follows: (1) The medical records and the rights of access are controlled by each EHR system, which makes the system more security. (2) Each transaction is linked to the previous transaction of the patient. It forms a chronological chain of hashes of medical records for a patient at all CDOs. This makes patients and users conveniently to track and verify whether or not the records have been tampered with. (3) Storing access permissions and logs on the blockchain not only makes it easy for patients to control who can be granted right to access their medical records, but also allows patients to know if the hospital is complying with the privacy policy. (4) SPV chain is more like today's healthcare system. It can be easily implemented using an existing public blockchain platform such as Ethereum and Hyperledger Fabric.

While, since the medical records are actually stored in each EHR system, users who want to access the medical records need to respectively send request to each EHR system, thus the EHR sharing process is less efficient. Additional security and privacy techniques such as encryption algorithms and access control need to be deployed in each EHR system to enhance the security and privacy of this medical data sharing model.

\subsubsection{Enhanced Techniques of SPV Chain}
\label{subsubsec:tech-SPV}

\begin{table*}[!t]%
\caption{Comparison of DSS Chain, STT Chain and SPV Chain}
\label{tab:compare}
\begin{minipage}{\columnwidth}
\begin{scriptsize}
\begin{center}
\begin{tabular}{|p{0.5cm}|p{1cm}|p{1.4cm}|p{3cm}|p{2.7cm}|p{2.4cm}|p{2.3cm}|p{1.8cm}|}
\toprule
 \makecell[l]{Chain}&\makecell[l]{Features}&\makecell[c]{Types}&\makecell[c]{Properties}&\makecell[c]{Advantages} &\makecell[c]{Disadvantages} &\makecell[c]{Needed security and\\ privacy techniques}&\makecell[c]{Typical systems}\\\hline
\makecell[l]{DSS}&\makecell[l]{Plaintext \\on chain}&\makecell[c]{Private\\ Consortium}&\makecell[c]{Consistency\\ Tamper-resistance\\Resistance to DDoS attacks\\Pseudonymity}&\makecell[l]{Records can be efficiently\\ shared. It is easily to track \\ the patient's records.}&\makecell[l]{Records are leaked \\to users who do not\\ need to access them.\\ It requires a trusted \\ party to authenticate\\ users.}&\makecell[l]{Anonymous\\ authentication,\\ anonymous signature,\\ security analysis \\ and verification of\\ smart contracts}&\\\hline
\makecell[l]{STT}&\makecell[l]{Ciphertext\\ on chain}&\makecell[l]{Consortium}&\makecell[c]{Consistency \\Tamper-resistance\\ Resistance to DDoS attacks\\ Pseudonymity\\ Confidentiality\\ Fine-grained access control\\ Tracking and auditing}&\makecell[l]{It supports fine-grained \\ access control and more\\ functions by using\\ different asymmetric\\ encryption algorithms.\\ It is easily to track the \\ transmission of records.}&\makecell[l]{It needs more storage\\ space and computation\\ resources, and is \\ vulnerable to collusion\\ attacks.}&\makecell[l]{ABE, HE, PRE, \\ NIZK, MPC, \\ security analysis and \\ verification of smart \\ contracts}&\makecell[l]{Medicalchain~\cite{Medicalchain}\\ MediBchain~\cite{Abdullah:2017:MediBchain}\\ DMMS~\cite{Patrick:2019:DMMS}}\\\hline
\makecell[l]{SPV}&\makecell[l]{Hash \\on chain}&\makecell[c]{Public}&\makecell[c]{Consistency\\ Tamper-resistance\\ Resistance to DDoS attacks\\ Patient-control\\ Tracking and auditing}&\makecell[l]{It is more secure than\\ other two architectures \\and easy to trace and\\ verify the records.\\ It can achieve public\\ auditing of access logs.}&\makecell[l]{Less efficient}&\makecell[l]{Security analysis and\\ verification of smart\\ contracts, TEE-based \\ smart contracts}&\makecell[l]{MedRec~\cite{Azaria:2016:MedRec}\\ FHIRChain~\cite{Zhang:2018:FHIRChain}\\ BlocHIE~\cite{Jiang:2018:BlocHIE}}\\
 \bottomrule
\end{tabular}
\end{center}
\end{scriptsize}
\bigskip\centering
\end{minipage}
\vspace{-9mm}
\end{table*}%

\textbf{Security Verification of Smart Contracts:}
MedRec~\cite{Azaria:2016:MedRec} utilizes Ethereum's smart contracts to create intelligent representations of existing medical records that are stored within individual nodes on the network and constructs the contracts to contain metadata about the record ownership, permissions and data integrity. Since everyone can deploy and execute smart contracts on Ethereum, and once released the code cannot be modified, a vulnerable smart contract will lead to information disclosure and even put the entire system at risk of being attacked. Therefore, an automated security analysis and verification program for smart contracts is necessary to prove contract behaviors as safe/unsafe with respect to a given property.

Fortunately, several research groups make sustained efforts to security analysis and validation of smart contracts and have developed automated analysis tools.
The symbolic execution tool, called Oyente~\cite{Luu:2016:MSC} built by Luu et al., which can detect vulnerabilities by extracting the control flow graph from the contract's EVM bytecode and executing it symbolically.
Bhargavan et al.~\cite{Bhargavan:2016:FVS} presented a platform that verifies various attributes on the code of functional language F*~\cite{Swamy:2016:DTM}, which is converted from the smart contract (i.e., Solidity or EVM bytecode).
Tsankov et al. developed a lightweight verifier for Ethereum smart contracts and named it Securify~\cite{Tsankov:2018:SPS}. Securify leverages the domain-specific insight that violations of many practical properties for smart contracts also violate simpler properties, which are significantly easier to check in a purely automated way.
Abdellatif et al.~\cite{Abdellatif:2018:FVSC} proposed a approach to model smart contract and blockchain execution protocol along with users' behaviors based on a formal model checking language. Based on these model implementations, and given their expected behavior, design vulnerabilities of the smart contracts can be analyzed using a statistical model checking tool.
Kalra et al. presented the design and implementation of ZEUS~\cite{Kalra:2018:ZEUS} - a practical framework for automatic formal verification of smart contracts using abstract interpretation and symbolic model checking. ZEUS takes as input the smart contracts written in high-level languages and leverages user assistance to help generate the correctness and/or fairness criteria in a XACML styled template.

\noindent \textbf{The TEE based Smart Contracts:}
The trusted execution environment (abbreviated as TEE) implies that an execution environment can provide an absolutely isolated application execution environment, so that it effectively protects application running in it from being tampered with and peeking at its operating state by other software applications and operating systems.
A representative implementation of TEE is Intel Software Protection Extensions (SGX) technology, which allows developers to partition sensitive information into enclaves with more security protection.

In recent years, researchers have tried to leverage SGX to solve the privacy protection problem in the implementation of smart contracts in the blockchain.
In 2018, Cheng and his co-authors proposed a confidentiality-preserving platform called Ekiden~\cite{Raymond:2018:Ekiden} for smart contracts based on Intel's SGX technology. The basic idea of Ekiden is to separate computation from consensus in a blockchain system. More specifically, the computation of the smart contract is performed in the TEEs of the off-chain compute nodes and a remote attestation protocol is used to verify the correctness of the computation. While the consensus nodes that are not equipped with trusted hardware are only used to maintain the normal operation of the blockchain system.
Moreover, Enigma~\cite{Zyskind:2015:Enigma} uses TEE in its current improved version, supporting users to develop privacy smart contracts.

\subsection{Comparison}
\label{subsec:compare}
We summarize the pros and cons of the three reference healthcare blockchain scenarios in Table~\ref{tab:compare}. Depending on the data stored on the blockchain, the blockchain plays a different role in each medical data sharing architecture. In DSS chain, the plaintexts of medical records are stored directly on the chain, and the blockchain is a data sharing platform. In STT chain, the ciphertexts of medical records encrypted by the receivers' private keys are stored on the chain, and the blockchain can be seen as an efficient and secure transfer platform. In SPV chain, only the hashes of medical records are stored on the chain, and blockchain is a security technology that helps EHR systems achieve secure storage and auditing of medical records.

Since the blockchain stores different content in the three chains, the security and privacy technologies additional required are different with each other. Note that, our focus here is on the security and privacy technologies that are needed in the blockchain. In the SPV chain, the EHR systems of course also need security and privacy technologies such as encryption and access control, but we only pay attention to the security and privacy technologies required by the blockchain, so we only listed security analysis and verification of smart contracts and TEE based smart contracts in Table~\ref{tab:compare}. Additional security and privacy techniques are required for all three architectures such as security analysis and verification of smart contracts and TEE based smart contracts, and some are uniquely required for one of the tree architectures such as asymmetric encryption algorithms are easier to be implemented and deployed in the STT chain.

\section{Other Healthcare Usage Scenarios}
\label{sec:other-usage-scenarios}
In addition to the EHR sharing, the blockchain can also be used in other healthcare application scenarios, such as federated learning for healthcare research, healthcare payment and drug tracking and medication auditing.

\subsection{Privacy-Preserving Federated Learning}
\label{subsec:blockchain-DL}
Currently, various healthcare research institutions use deep learning technique for healthcare research over their own medical data. The data held by each institute is limited and has certain regional characteristics. If researchers want more general results, they need to share medical data with researchers in other institutes, which will lead to the leakage of private information. Federated learning, a.k.a. collaborative learning, is a solution for the combination of deep learning and distributed computation, where there is a parameter server, maintaining a deep learning model to train and multiple parties that take part in the distributed training process with their own data. But federated learning cannot provide the privacy for the training data, even the training data is stored and trained locally.

Jiasi Weng et al.~\cite{Weng:2018:DeepChain} proposed a privacy-preserving blockchain-based federated learning framework named DeepChain, where the blockchain is used as a medium for transmitting local intermediate gradients and parameters between each training model and the workers (who act as a server to compute parameters on the collected gradients). DeepChain uses paillier cryptosystem~\cite{Nishide:2011:DPC} to encrypt the gradients that sent to blockchain network, so that works can calculate the parameters over the encrypted data without leaking any information about local gradients. The gradients and parameters can be easily verified whether they are generated correctly with the cryptographic primitives. It also uses an incentive mechanism to encourage parties to participate in the federated learning and behave honestly.

\subsection{Healthcare Payments}
\label{subsec:blockchain-pay}
Although healthcare payments between healthcare providers, payers and patients have been converted from paper to electronic systems, they are less efficient and prone to disputes.
Patients are often confused by the inconsistent statements of healthcare providers and payers. Patients also have to provide identity information to register and log in the EHR systems of different CDOs to check the their medical data and statements of healthcare provider and payers.
The payer wants to confirm that the healthcare expenses are consistent with the healthcare services provided by the healthcare provider, and there are no unreasonable expenses. They also want to pay the healthcare service providers more easily and frictionless, but they usually do not know whether the patients have paid the provider.
Providers want to avoid sending multiple statements about healthcare services to payers and patients. They also hope that healthcare expenses can be paid quickly and without friction.

Recently, InstaMed built a private blockchain prototype based on HyperLedger Fabric 1.4 to provide efficient and convenient healthcare payments among healthcare providers, payers and patients~\cite{InstaMed}.
The main goal of developing such a prototype is to leverage the advantages of blockchain to eliminate inconsistencies in statements and friction for all stakeholders so that they can get a better payment experience.

\subsection{Drug Tracking and Medication Auditing}
\label{subsec:blockchain-drug}
Taking fake drugs or drugs of unknown origin not only cannot achieve the effect of treatment, but also affects people's health, and even leads to death. According to a report of Health Research Funding Organization (HRFO), hundreds of people around the world die due to the drug counterfeiting each year.
Thus, it is important to track the whole life cycle of drugs from production to use.

Blockchain's distributed storage and tamper-resistant properties make it ideal for drug tracking. Any particular drug registered on the blockchain should be proven to be authentic. Specifically, pharmaceutical companies can store their certificates and approval certificates and clinical trial results of drugs on blockchain to prove that the drugs produced by them are authentic. In addition, the transparent feature of the data on the blockchain helps to find the complete path of origin, thereby helping to eradicate the circulation of counterfeit medicines.
In 2017, Chronicled launched a project called MediLedger~\cite{MediLedger} to innovate evolving solutions for drug track and trace regulations, and provide a step function improvement in the overall operation of the drug supply chain. It brings together leading pharmaceutical manufacturers and distributors using an advanced and customizable decentralized supply chain management system based on the principles of blockchain. The MediLedger Project is building an industry-owned permissioned blockchain network for the pharmaceutical sector based on open standards and specifications.

Another problem that can be solved with the tamper-resistance features of the blockchain is the prescription fraud, including modifying numbers and content of prescription, duplication of prescriptions, and so-called ``doctor shopping" whereby fraudsters visit many doctors to collect as many as original prescriptions as possible~\cite{Engelhardt:2017:HHC}.
Nuco (Aion)~\cite{Nuco:wiki}, which was started in 2016, is working on blockchain solutions to combat prescription fraud. Nuco uses permissioned blockchain to solve the incomplete feedback between the prescription writers (physicians) and the prescription fillers (pharmacists).

\begin{table*}%
\caption{State-of-the-art Healthcare Blockchains}
\label{tab:related-work}
\begin{minipage}{\columnwidth}
\begin{footnotesize}
\begin{center}
\begin{threeparttable}
\begin{tabular}{|p{1.5cm}|p{2cm}|p{0.5cm}|p{6.2cm}|p{0.2cm}p{0.2cm}p{0.2cm}p{0.2cm}p{0.2cm}p{0.2cm}p{0.2cm}p{0.2cm}p{0.2cm}p{0.2cm}|}
\toprule
\makecell[c]{Types}&\makecell[c]{Authors}&Year&\makecell[c]{Key objective}&1&2&3&4&5&6&7&8&9&10\\\hline
\multirow{4}{1.5cm}{}&\makecell[l]{Omar et al.~\cite{Abdullah:2017:MediBchain}} & 2017 & To develop a privacy-preserving and patient centric healthcare data management platform based on blockchain. & \checkmark & \checkmark & \checkmark & \checkmark & \checkmark & \checkmark & \checkmark & \checkmark & \XSolid & \checkmark\\
 &\makecell[l]{Xia et al.~\cite{Xia:2017:BBDS}}& 2017 & To design a blockchain-based data sharing framework to achieve fine-grained access control for sensitive data outsourced in the cloud. & \checkmark & \checkmark & \checkmark & \checkmark & \checkmark & \XSolid &  \XSolid &  \XSolid & \checkmark & \checkmark\\
 &\makecell[l]{Xia et al.~\cite{Xia:2017:MeDShare}}& 2017 & To develop a system called MeDShare for medical data sharing among medical big data custodians in a trust-less environment. & \checkmark & \checkmark & \checkmark & \checkmark & \checkmark & \XSolid &  \XSolid &  \XSolid & \checkmark & \checkmark\\
 &\makecell[l]{Liang et al.~\cite{Liang:2017:IBA}} & 2018 & To propose an user-centric health data sharing solution based on a permissioned blockchain to achieve privacy-preserving.& \checkmark & \checkmark & \checkmark & \checkmark & \checkmark & \XSolid & \checkmark & \XSolid & \checkmark & \checkmark\\
&\makecell[l]{Fan et al.~\cite{Fan:2018:MedBlock}} & 2018 & To develop a blockchain-based medical data management system, MedBlock, to handle medical data in a way of privacy protection.& \checkmark & \checkmark & \checkmark & \XSolid & \checkmark & \checkmark & \checkmark & \checkmark & \checkmark & \checkmark\\
\makecell[c] {Consortium} &\makecell[l]{Zhang and Lin\\~\cite{Zhang:2018:TSP}} & 2018 & To propose a blockchain-based secure and privacy-preserving personal health information sharing scheme. &\checkmark & \checkmark & \checkmark & \checkmark & \checkmark & \checkmark & \checkmark & \checkmark & \checkmark & \checkmark\\
 &\makecell[l]{Li et al.~\cite{Patrick:2019:DMMS}}& 2019 & To design a decentralized medication management system (DMMS) based on the hyperledger fabric framework to manage medication histories. & \checkmark & \checkmark & \checkmark & \checkmark & \checkmark & \XSolid & \checkmark & \checkmark & \checkmark & \checkmark\\
 &\makecell[l]{Choudhury et al.\\~\cite{Choudhury:2019:ABF}} & 2019 & To propose a data management framework based on permissioned blockchain with smart contracts and private channels. &\checkmark & \checkmark & \checkmark & \checkmark & \checkmark & \checkmark & \XSolid & \checkmark & \checkmark & \checkmark\\
 &\makecell[l]{Tanwar et al.~\cite{Tanwar:2020:Blockchain-based}} & 2020 & To propose a blockchain-based approach for implementing a permission-based EHR sharing system with symmetric key encryption. &\checkmark & \checkmark & \checkmark & \checkmark & \checkmark & \XSolid & \checkmark & \checkmark & \checkmark & \checkmark\\
 &\makecell[l]{Huang et al.~\cite{Huang:2020:BBS}} & 2020 & To propose a blockchain-based privacy-preserving scheme to realize secure sharing of medical data among different entities. & \checkmark & \checkmark & \checkmark & \checkmark & \checkmark & \checkmark & \checkmark & \checkmark & \checkmark & \checkmark\\\hline
 \multirow{4}{1.5cm}{}&\makecell[l]{Azaria et al.~\cite{Azaria:2016:MedRec}} & 2016 & To develop a blockchain-based EMR sharing system with smart contracts. & \checkmark & \checkmark & \checkmark & \checkmark & \checkmark & \checkmark & \checkmark & \XSolid & \checkmark & \checkmark \\
 &\makecell[l]{Yang et al.~\cite{Yang:2017:ABB}} & 2017 & To propose a blockchain-based approach by combining signcryption and attribute-based authentication to realize secure healthcare data sharing.& \checkmark & \checkmark & \checkmark & \checkmark & \checkmark & \checkmark & \checkmark & \checkmark & \checkmark & \checkmark\\
 &\makecell[l]{Zhang et al.~\cite{Zhang:2018:FHIRChain}} & 2018 & To realize FHIRChain, a blockchain-based architecture, to achieve efficient clinical data sharing. & \checkmark & \checkmark & \checkmark & \checkmark & \checkmark & \checkmark & \XSolid & \checkmark & \checkmark & \checkmark \\
 \makecell[c]{Public}&\makecell[l]{Jiang et al.~\cite{Jiang:2018:BlocHIE}} & 2018 & To realize BlocHIE, a blockchain-based platform for healthcare data exchange. & \checkmark & \checkmark & \checkmark & \XSolid & \checkmark & \XSolid & \XSolid & \XSolid & \XSolid & \XSolid \\
 &\makecell[l]{Li et al.~\cite{Li:2018:BBD}} & 2018 & To design a blockchain-based data preservation system (DPS) for medical data and use cryptographic algorithms to achieve user privacy. & \checkmark & \checkmark & \checkmark & \XSolid & \checkmark & \checkmark & \checkmark & \checkmark & \checkmark & \XSolid\\
 &\makecell[l]{Zou et al.~\cite{Zou:2021:SPChain}} & 2021 & To develop a public blockchain-based medical data sharing and privacy-preserving eHealth system, SPChain, with high throughput. &\checkmark & \checkmark & \checkmark & \checkmark & \checkmark & \XSolid & \checkmark & \checkmark & \checkmark & \checkmark\\
\bottomrule
\end{tabular}
\begin{tablenotes}
\item 1: Consistency of transactions 2: Integrity of transactions 3: Authenticity of transactions 4: Tracking and auditing of access 5: Availability of system and transactions 6: Anonymity of users 7: Patient-control 8: Confidentiality of transactions 9: Fine-grained access control 10: Authentication of users.
\end{tablenotes}
\end{threeparttable}
\end{center}
\end{footnotesize}
\bigskip\centering
\end{minipage}
\vspace{-9mm}
\end{table*}%

\section{Comparison to other related Work}
\label{sec:state-of-the-art}
We have described the security and privacy issues and challenges in healthcare blockchains from the following ten perspectives: (1) Consistency of transactions; (2) Integrity of transactions; (3) Authenticity of transactions; (4) Tracking and auditing of access; (5) Availability of system and transactions; (6) Anonymity of users; (7) Patient-control; (8) Confidentiality of transactions; (9) Fine-grained access control; and (10) Authentication of users. Table 3 provides a summary of some representative works in the literature from these ten perspectives.

In short, academia, industry, government agencies, and medical organizations have engaged in intense discussions and explorations on the application of blockchain technology in the healthcare sector since 2016.

In academia, discussions on the application of blockchain technology to medical information systems have continued for several years, and most scholars have given positive answers and raised challenges~\cite{Deloitte:2016:Blockchain,Ramachandran:2020:Review,Lemieux:2021:Having,Ibrar:2021:Blockchain}.
Asaph Azaria and his colleagues proposed MedRec~\cite{Azaria:2016:MedRec} in 2016, a decentralized record management system using the blockchain technology. It uses an immutable log for patients to access their EMR data across multiple care providers and medical treatment facilities, providing authentication, confidentiality, accountability for data sharing.
Inspired by the work of MedRec~\cite{Azaria:2016:MedRec}, Kevin Peterson and his colleagues~\cite{Peterson:2016:ABB} presented a blockchain-based approach to sharing patient data. Instead of a single centralized source of trust, it uses decentralized network consensus and predicates consensus on proof of structural and semantic interoperability.
Then, several blockchain-based healthcare data management and sharing systems~\cite{Yang:2017:ABB,Abdullah:2017:MediBchain,Xia:2017:BBDS,Xia:2017:MeDShare, Fan:2018:MedBlock,Zhang:2018:FHIRChain,Jiang:2018:BlocHIE,Zhang:2018:TSP,Li:2018:BBD,Patrick:2019:DMMS,Choudhury:2019:ABF,Tanwar:2020:Blockchain-based,Huang:2020:BBS,Zou:2021:SPChain} were proposed. The detailed comparison of these work is listed in Table 3.

In industry, some startups are committed to developing blockchain-based medical record storage and sharing systems that will revolutionize traditional medical record management.
Healthchain~\cite{Healthchain:GitHub} and ScalaMed~\cite{ScalaMed} have expressed a strong desire and emphasis on giving patients control over their data including the ability to authorize who can use it and how.
Medicalchain~\cite{Medicalchain} is a decentralized platform that enables secure, fast and transparent exchange and usage of medical data.

Government agencies and medical organizations are also actively exploring the application of blockchain in medical data exchange and refinement of security and privacy rules. Healthcare Information and Management Systems Society (HIMSS) published a guide~\cite{HIMSS} for better understanding the basic technologies and potential applications of blockchain in healthcare. In United Kingdom, Reform, a London-based organization, published a report~\cite{Reform} detailing the benefits and challenges the blockchain technology can bring to the U.K. through the National Health Service (NHS).One of the key challenges is complying with the General Data Protection Regulation (GDPR). China has taken an assertive and proactive stance. Alibaba Health announced the cooperation with the city of Changzhou on a Blockchain Pilot Project: applying blockchain technology to the underlying technical architecture system of the Changzhou Medical Consortium to achieve safe and controllable data interconnection between some local medical institutions~\cite{William}.

Another comparable blockchain system is the Sovrin~\cite{Sovrin:2018}, which is a blockchain based protocol for self-sovereign identity (SSI) in distributed network. It is designed to bring the distributed trust, personal control, and ease-of-use of digital IDs to the Internet without the intervening administrative authorities. Sovrin also uses pairwise pseudonyms and zero-knowledge proof to implement privacy for its global users. We believe that Sovrin (or SSI technique) is a potential solution that can be used in the next-generation healthcare blockchain to manage user identities securely and privately.

\section{Conclusion}
We have reviewed the security and privacy requirements, risks and solution techniques in healthcare blockchain for electronic medical data sharing, covering the following four aspects of the security and privacy in healthcare blockchain.
First, we discussed the security and privacy requirements of a healthcare blockchain.
Second, we categorized existing efforts into three reference blockchain usage scenarios based on the different forms of medical data stored on the blockchain and presented potentially available security techniques to enhance these three representative blockchain-based EHR sharing systems.
Third, we outlined other potential and attractive blockchain application scenarios in the healthcare sector.
We conjecture that this survey will help healthcare professionals, healthcare service developers and healthcare consumers to gain an in-depth and comprehensive understanding of the security and privacy requirements and technologies for enabling a secure and decentralized EMR data sharing through healthcare blockchain.

\section*{Acknowledgment}
Rui Xue acknowledge the support from National Natural Science Foundation of China under Grant No.: 61772514 and Beijing Municipal Science \& Technology Commission (No.: Z191100007119006). Ling Liu acknowledges a partial support from the USA National Science Foundation CISE grants 2038029, 2026945, and 1564097.

\ifCLASSOPTIONcaptionsoff
  \newpage
\fi

\bibliographystyle{IEEEtran}
\bibliography{references}
\end{document}